\documentclass[ijoc,nonblindrev]{informsArXiv} 

\OneAndAHalfSpacedXI

\usepackage{natbib}
 \bibpunct[, ]{(}{)}{,}{a}{}{,}%
 \def\bibfont{\small}%
 \def\BIBand{and}%

\TheoremsNumberedThrough     

\EquationsNumberedThrough    

\usepackage{graphicx}
\usepackage{algorithm}
\usepackage[noend]{algorithmic}
\algsetup{indent=2em}
\usepackage{amsmath}
\usepackage{mathrsfs} 
\usepackage{amsfonts}
\usepackage{amssymb}
\usepackage{boites}
\usepackage{natbib}
\usepackage{xcolor}

\newcommand{\var}{\mathbb{V}\mathrm{ar}} 
\newcommand{\cov}{\mathbb{C}\mathrm{ov}} 

\newcommand{\m}[1]{\mathbf{#1} }
\newcommand{\di}{\mathrm{d} }
\renewcommand{\v}[1]{\boldsymbol{#1}}
\newcommand{\bb}[1]{\mathbb{#1}}
\renewcommand{\c}[1]{\mathcal{#1}}
\newcommand{\s}[1]{\mathscr{#1}}
\newcommand{\idef}{:=}

\newcommand{\gvn}{\,|\,}

\newcommand{\EGS}{\mathbb{E}_\mathrm{GS}}

\def\EE{\mathbb{E}}
\def\II{\mathbb{I}}

\def\PP{\mathbb{P}}
\def\QQ{\mathbb{Q}}
\def\RR{\mathbb{R}}

\begin{document}


\RUNAUTHOR{Botev and L'Ecuyer}

\RUNTITLE{Sampling via Splitting}

\TITLE{Sampling Conditionally on a Rare Event via Generalized Splitting}

\ARTICLEAUTHORS{%
\AUTHOR{Zdravko I. Botev}
\AFF{UNSW Sydney, Australia, \EMAIL{botev@unsw.edu.au}}
\AUTHOR{Pierre L'Ecuyer}
\AFF{Universit\'e de Montr\'eal, Canada, and Inria--Rennes, France \EMAIL{lecuyer@iro.umontreal.ca}}
} 

\ABSTRACT{We propose and analyze a generalized splitting method to sample approximately from
a distribution conditional on the occurrence of a rare event.
This has important applications in a variety of contexts in operations research, 
engineering, and computational statistics.  
The method uses independent trials starting from a single particle.
We exploit this independence to obtain asymptotic and non-asymptotic bounds on 
the total variation error of the sampler. 
Our  main finding is that the approximation error  depends crucially on the relative variability 
of the number of points produced by the splitting algorithm in one run, 
and that this relative variability can be readily estimated via simulation. 
We illustrate the relevance of the proposed method on an application in which one needs to
sample (approximately) from an intractable posterior density in Bayesian inference.
}


\maketitle

\section{Introduction}

We consider the problem of generating samples from a conditional distribution when the 
conditioning is on the occurrence of an event that has a small probability.
We have a random variable $\v X$ defined over
a probability space $(\RR^d, \c B, \PP)$, where $\c B$ can be taken as the Borel sigma-field,
and $\v X$ has a probability density function (pdf) $f$. 
We assume it is easy to sample exactly from the density $f$.
The rare event on which we condition can be written in the form $B = \{S(\v x)\geq\gamma\} \in\c B$
for an appropriately chosen measurable function $S : \RR^d\to\RR$ called the \emph{importance function}.
The conditional pdf is then
\begin{equation}
\label{eq:cond}
  q(\v x) = \frac{f(\v x)\II\{S(\v x)\geq\gamma\}}{\ell(\gamma)} , \quad \v x=(x_1,\ldots,x_d)^\top,
\end{equation}
where $\II$ is the indicator function, and 
\begin{equation}
\label{eq:ell}
  \ell = \ell(\gamma)=\bb P(S(\v X)\geq\gamma)
\end{equation} 
is the appropriate (unknown) normalizing constant, 
which we assume is so small that estimating it via the naive acceptance-rejection method 
(simulate $\v X\sim f$ until $S(\v X)\geq\gamma$) is impractical.

Sampling from a distribution conditional on a rare event has many applications.
For example, suppose we want to generate $\v X$ from an arbitrary density 
proportional to $p(\v x)$ for $\v x\in \bb R^d$, for some known function $p$,
and that it is too hard to generate samples directly from this density.  
Since $p$ is known, we may be able to find a density $f$ such that
$\sup_{\v x} {p(\v x)}/{f(\v x)} < \gamma$ for some constant $\gamma<\infty$. 
Then to generate $\v X$, it suffices to generate a pair of independent random variables 
$\v X\sim f$ and $U\sim \mathsf{U}(0,1)$ 
conditional on  the event ${p(\v X)}/{f(\v X)}\geq \gamma U$, which is frequently a rare event \citep{kroese2013handbook}[Section 14.5].
This fits our framework by taking $S((\v x,u)) = p(\v x)/(f(\v x) u)$.

Another application is Bayesian Lasso regression \citep{park2008bayesian}, 
 in which inference requires 
repeated simulation of a vector $\v \beta$ of model parameters, 
conditional on the regularization constraint $\|\v\beta\|_1<\gamma$.
We give a detailed example of this in Section~\ref{sec:example}.

A third type of application occurs in the setting where we want to estimate the probability $\ell$ of the rare event and to understand under which circumstances the rare event
is likely to occur.
A popular method to estimate $\ell$ is importance sampling, and the optimal way to do it 
is to sample under a density $f$ proportional to the original density conditional 
on the rare event, and then adjust the estimator using a likelihood ratio 
\citep{vTUF14a,botev2014variance,botev2011importance}.
This also fits our framework.
In this context, it can be very useful to sample from the conditional density to 
get insight on how the rare event occurs.  
For instance, in a network with unreliable links, one may want to sample random configurations
of all the links conditionally on a failure of the network, 
to better understand what (typically) makes the network fail \citep{botev2014reliability,botev2012dependent}.

The sampling methods examined in this paper are based on 
the \emph{generalized splitting} (GS) algorithm of \cite{botev2012efficient} for drawing a collection of 
random vectors whose distribution converges to a target distribution with pdf of the form \eqref{eq:cond}.
To apply GS, we first select an increasing sequence of levels 
$-\infty = \gamma_0 < \gamma_1 < \cdots < \gamma_\tau = \gamma$ for the importance function $S$.
This can be done in pilot runs via a run \citep{botev2012efficient}.
The algorithm uses a branching process that favors states $\v X$ having a large value of $S(\v X)$ 
by resampling them conditional
on staying above the current threshold, thus ``splitting'' those states into new copies, 
and then discarding those that do not reach the next level. 
At the end, the states that have reached the last level $\gamma$ are retained.
This process is replicated several times independently and all the retained states are collected
to form an empirical version of the target conditional distribution.
There are many ways of choosing the total number  of replications (or \emph{trials}).
For example, one can fix them in advance to a constant $n$, or one can repeat the procedure until 
$n$ trials have provided at least one retained state each,
or until the total number of retained states is more than $t$, 
or until a certain computing budget (CPU time) has elapsed.
In the latter case, one can either complete the current trial, or discard it, or just take the states
retained so far from that trial.

There is a large variety of splitting-type or interacting particle algorithms 
to sample the state of a Markov chain approximately from its steady-state
distribution conditional on a rare event; see for example
\cite{vGLA99d,vCER05a,cerou2012sequential,vLEC09b,tAND10a,vBRE16a}, and the references given there.
The analysis of these algorithms consists in most cases in proving their unbiasedness
when estimating the expectation of a random variable 
that can be nonzero only
when the rare event occurs (estimating the probability of the rare event is a
special case of this), and sometimes showing their asymptotic efficiency when 
the probability of the rare event decreases toward zero \citep{vDEA09a}.

In this paper, we are interested in the different problem of bounding the 
difference between the exact conditional distribution and the distribution obtained by
picking a state from the sample returned by the splitting algorithm.
We do this for some variants of the GS method of \cite{botev2012efficient}.
\cite{vLEC18s} proved that this method provides an unbiased estimator of the expected 
value of a cost function, but also showed that a state picked at random from the set
of retained states at the last level does not follow the true conditional  distribution 
in general. 

 On the other hand, the distance between the two distributions converges to
zero when the number of replicates increases toward infinity.
The aim of the present paper is to study how this convergence occurs and to 
establish explicit \emph{non-asymptotic} (risk) bounds on the total variation (TV) error between the two distributions, their mean absolute value,
and the expectation of the TV error in the case when it is a random variable. Our bounds are expressed in terms of simple mathematical expectations that can be estimated 
easily from the simulation output. 

We provide convergence results for two versions of the GS algorithm. 
In both, we assume that whenever a trial returns no state from the rare event set (an empty trial), 
we discard it and try again.  
In the first version, we run GS until we have $n$ non-empty trials, for some fixed $n > 0$.
We prove that the TV distance between the true conditional distribution and the 
distribution of a state picked at random from the retained states from this GS version is 
bounded by $c_1/n$ where $c_1$ is an unknown constant that can be estimated from the simulation output.
In the second version, we run GS until the total number of retained states exceeds $t$, for some fixed positive integer $t$. 
For this version, we show that the convergence rate is of the form $c_2(t) t^{-3/2}=\c O(t^{-3/2})$, where the quantity $c_2(t)$ is bounded uniformly in $t$ and can be estimated from the simulation output.
The derivation of these bounds is made possible thanks to the fact that GS produces independent trials,
each one starting from a single particle, and this permits us to use results from renewal theory for our analysis.

Typically, approximate simulation from the target pdf  \eqref{eq:cond} is accomplished 
using Markov chain Monte Carlo (MCMC) (for example, \cite{jones2001honest,taimre2019MCM}). 
While MCMC sampling can be  simple to implement, it still poses the challenge of analyzing its output and deciding how close the 
sampling or empirical distribution is to the desired target distribution \citep{jones2001honest}. The reason for this 
difficulty is that  MCMC  generates  a sequence of \emph{dependent} random vectors $\v Y_1,\v Y_2,\ldots$. 
Typical graphical diagnostic tools like autocorrelation plots are  heuristics, which do not easily 
provide precise qualitative measure of how close the simulated random variables follow  the target distribution. 
Also we have to choose from an infinite number of possible one-dimensional plots.
In contrast, our bounds on the TV error present a more rigorous and theoretically justified 
convergence assessment than the autocorrelation plots typically used in MCMC.

The rest of the paper is organized as follows. 
In Section~\ref{sec:gs}, we recall the GS algorithm used in this paper.  
In Section~\ref{sec:cond-sampling}, we define our versions of GS used for sampling conditional on the rare event.  
In Section~\ref{sec:convergence}, we state our main new results on the convergence of the distance between 
the empirical and true conditional distributions, and bounds on this distance.
The proofs are given in the appendix.
In Section~\ref{sec:example}, we show how our methodology can be applied in a practical setting, 
namely to sample approximately from the posterior density of the Bayesian Lasso.
In this example, we show how the non-asymptotic risk bounds can be used to assess convergence and to 
estimate the error committed when using GS to sample from the conditional distribution. We also compare the simulation accuracy of GS with that of the sequential Monte Carlo method  \citep{cerou2012sequential}.

\section{Background on Generalized Splitting}
\label{sec:gs}

We recall the GS method for estimating the rare-event 
probability $\ell$ in \eqref{eq:ell}. 
This method is a simple generalization of the classical multilevel splitting technique
for rare-event simulation 
\citep{kahn1951estimation,vGLA99d,garvels2002importance,vLEC09b}. Our background material
 here is similar to  the one given in \cite{vLEC18s}.

The  idea of GS is to define a discrete-time Markov chain with state $\v Y$, which evolves via a  branching-type random mechanism that pushes it toward 
a state corresponding to  $\{\v Y\in B\}\equiv\{S(\v Y)\geq\gamma\}$ in \eqref{eq:cond}. 
To estimate  the (rare-event) probability \eqref{eq:ell} via GS, we first need to choose: 
\begin{enumerate}
\item an integer $s \ge 2$, called the \emph{splitting factor}, and 
\item an integer $\tau > 0$ and real numbers 
$-\infty=\gamma_0 < \gamma_1 < \dots < \gamma_{\tau} = \gamma$ for which 
\begin{equation*}
 \rho_l \idef \bb P(S(\v Y) \geq \gamma_l \mid S(\v Y) \geq \gamma_{l-1}) \approx 1/s
\label{eq:rhot}
\end{equation*}
for $l=1,\dots,\tau$ (except for $\rho_{\tau}$, which can be larger than $1/s$).
These $\gamma_l$'s represent the $\tau$ levels of the splitting algorithm.  In Section~\ref{sec:example} we give particular choices of $s$ and $\tau$ that are relevant to our examples.
\end{enumerate}

For each level $\gamma_l$ we construct a Markov chain whose stationary density is equal to
the density of $\v Y$ conditional on $S(\v Y) \geq \gamma_l$ (a truncated density), given by 
\begin{equation}
 q_l(\v y) \idef f(\v y) \frac{\bb I\{S(\v y) \geq \gamma_l\}}{\bb P(S(\v Y) \geq \gamma_l)}.
\label{eq:density-ft}
\end{equation}
Note that $q_0 = f$ and $q_{\tau} = q$.
We denote by $\kappa_l \equiv \kappa_l(\cdot\mid\cdot)$ the transition kernel of this Markov chain:
$\kappa_l(\di \v y \mid \v x)$ represents the probability that the next state is in $\di \v y$
when the current state is $\v x$.
There are many ways of constructing this Markov chain and $\kappa_l$.
A practical example using Gibbs sampling will be given in Section~\ref{sec:example}.


\begin{algorithm}[H]
\caption{GS Sampler 1}
\label{algo:splitting}
{\small
\begin{algorithmic}
\REQUIRE {$s$, $\tau$, $\gamma_1,\dots,\gamma_{\tau}$}
 \STATE{generate a vector $\v Y$ from its unconditional density $f$}
 \IF{$S(\v Y) < \gamma_1$}
   \RETURN {$\c X_{\tau} = \emptyset$ and $\mathring{M} = 0$}
 \ELSE
   \STATE{$\c X_1 \leftarrow \{\v Y\}$ \ \ }
	     	 \COMMENT{this state $\v Y$ has reached at least the first level}
   \FOR{$l=2$ \TO $\tau$}
     \STATE{$\c X_l \leftarrow \emptyset$ \ \ }
         \COMMENT{list of states that have reached the level $\gamma_t$}
     \FORALL {$\v Y \in\c X_{l-1}$}
       \STATE{set $\v Y_0 \leftarrow \v Y$ \ \ }
			    \COMMENT{we will simulate $s$ steps from this state}
       \FOR {$j=1$ \TO $s$}
         \STATE {sample $\v Y_{j}$ from  $\kappa_{l-1}(\di \v y\mid \v Y_{j-1})$}
         \IF{$S(\v Y_{j}) \geq \gamma_l$}
           \STATE{add $\v Y_{j}$ to $\c X_l$ \ \ }
					     \COMMENT{this state has reached the next level}
         \ENDIF 
       \ENDFOR
     \ENDFOR
   \ENDFOR
 \ENDIF 
 \RETURN {the list $\mathring{\c Y}=\c X_\tau$ of retained states and its cardinality $\mathring{M}$.}
\end{algorithmic}
}
\end{algorithm}

The original GS algorithm is summarized in Algorithm~\ref{algo:splitting}, and is also given in \cite{vLEC18s}. 
The algorithm returns a list $\mathring{\c Y}$ of  \emph{retained states} 
that belong to  $B=\{\v y: S(\v y)\geq \gamma\}$, as well as the size of this list. 
This list is a multiset, in the sense that it may contain the same state more than once. 
The list $\mathring{\c Y}$ can be empty and its cardinality $\mathring{M}=0$.
The o-ring symbol in the notation is a reminder that the size of the set can be zero. In the remainder of this article,  we define $\c Y$ and $M$ as the versions of $\mathring{\c Y}$ and $\mathring{M}$, 
conditional on $\mathring{M}\geq 1$.

Let $\c A $ denote a $\sigma$-algebra of Borel measurable subsets of $\bb R^d$.
For some of our results, $\c A$ will be a more restricted class 
than the Borel subsets of $\bb R^d$.
Algorithm~\ref{algo:splitting} can be used to  estimate $\bb P(\v Y \in A)$ for any $A\in\c A$ 
via the unbiased estimator:
\begin{equation}
\label{hatell}
\widehat{\bb P}(A)=\mathring{H}(A)/s^{\tau-1},
\end{equation}
where $\mathring{H}(A)=|\mathring{\c Y} \cap A|$ is  the number of states $\v Y\in \mathring{\c Y}$ 
that belong to $A$.  
In practice, one will replicate this algorithm several times and take the average.
The unbiasedness is implied by the following lemma, proved in \cite{vLEC18s}.

\begin{lemma}[\cite{vLEC18s}]
\label{thrm:expectation}
For any measurable function $\hbar : \mathbb{R}^d\mapsto \bb R$ and any measurable subset $A \subseteq B$, we have
\begin{equation}
\label{eq:expected-cost1}
  \EGS\left[\sum_{\v Y\in\mathring{\c Y}}  \hbar(\v Y) \bb I\{\v Y\in A\}\right]
    = s^{\tau-1} \bb E[\hbar(\v Y) \bb I\{\v Y\in A\}],
\end{equation}
where the  expectation on the left-hand-side is with respect to $\mathring{\c Y}$  from  Algorithm~\ref{algo:splitting}
and the  expectation on the right-hand-side is with respect to the original density $f$.
\end{lemma}

By taking $\hbar$ as the identity function in \eqref{eq:expected-cost1}, 
we obtain that $\widehat{\bb P}(A)$ in \eqref{hatell} is unbiased for $\PP(A)$:
\begin{equation*}\textstyle
\label{eq:expected-cost2}
  \EGS[\mathring{H}(A)] = \EGS\left[\sum_{\v Y\in\mathring{\c Y}} \bb I\{\v Y\in A\}\right] = s^{\tau-1} \bb P(\v Y \in A),
\end{equation*}
and therefore
\begin{equation*}
\label{eq:expected-cost3}
  \EGS[\widehat\PP(A)] = \EGS[\mathring{H}(A)/s^{\tau-1}] = \bb P(\v Y \in A).
\end{equation*}
Moreover, since  $\EGS[\mathring{M}] = s^{\tau -1} \bb P(\v Y \in B)$ and $A\subseteq B$, 
we have that 
\[
\frac{\EGS[\mathring{H}(A)]}{\EGS[\mathring{M}]} 
  = \bb P(\v Y \in A \mid \v Y \in B)
	= \frac{\bb P(\v Y \in A)}{\bb P(\v Y \in B)}.
\]

\section{Sampling Conditionally on a Rare Event}
\label{sec:cond-sampling}

When estimating an expectation as in (\ref{eq:expected-cost1}), an empty list $\mathring{\c Y}$
poses no problem: the unbiased estimator just takes the value 0 in that case.
But for our purpose of sampling from a conditional distribution, we insist that there are no  empty sets of retained states.  
To make sure that the set of retained states is non-empty, we modify the original 
GS so that each trial returns at least one state.  
Whenever a GS run returns an empty list, we simply discard it and try again.
This gives Algorithm~\ref{algo:splitting2}.

\begin{algorithm}[H]
\caption{ GS Sampler 2}
\label{algo:splitting2}
{\small
\begin{algorithmic}
\REQUIRE {$s$, $\tau$, $\gamma_1,\dots,\gamma_{\tau}$}
\REPEAT
  \STATE{run Algorithm~\ref{algo:splitting}}
\UNTIL {$\mathring{M} > 0$}
\RETURN {the list ${\c Y}=\c X_\tau$ of retained states and its cardinality ${M} = |\c Y|$.}
\end{algorithmic}
}
\end{algorithm}

Does this algorithm still provide an unbiased estimator?
An important observation is that if we replace $\mathring{\c Y}$ by $\c Y$ 
in (\ref{eq:expected-cost1}), the equality is no longer true.
That is, we get a \emph{biased} estimator of the expectation on the right.
However, our main goal here is not to estimate this expectation, but to sample
approximately from the conditional distribution, and we will analyze methods that use
Algorithm~\ref{algo:splitting2} for this purpose. 
As mentioned earlier, there are several ways of doing it. 
In this paper, we examine the following two versions:
(a) run a fixed number $n$ of iid replicates of Algorithm~\ref{algo:splitting2} and
(b) perform replicates until there are more than $t$ retained states in total. 
These two approaches are detailed in Algorithms~\ref{algo:n iid} and \ref{algo:until t}, respectively.
In both cases, at the end we collect all the retained states in a multiset $\c Y_{\cup}$.
%
For the first version the cardinality of the returned set $\c Y_{\cup}$ is at least $n$, 
whereas in the second case it is at least $t$ 
and $N(t)$ is the (random) number of calls to Algorithm~\ref{algo:splitting2}. We summarize these two versions as follows.

\begin{algorithm}[H]
\caption{ Sampling an empirical distribution from $n$ iid non-empty GS replications}
\label{algo:n iid}
{\small
\begin{algorithmic}
\REQUIRE {$s, \tau, \gamma_1,\dots,\gamma_{\tau}$ and $n$}
\FOR{$i=1,\ldots,n$}
  \STATE{run  Algorithm~\ref{algo:splitting2} to obtain the list ${\c Y}_i$ of size $M_i$}
\ENDFOR
\RETURN{the empirical distribution $\widehat{\bb Q}_n$ of  the states in the set $\c Y_\cup:=\c Y_1\cup\cdots\cup\c Y_n$}
\end{algorithmic}
}
\end{algorithm}

\vspace{-.8cm}
\begin{algorithm}[H]
\caption{Sampling an empirical distribution with  more than $t$ retained states}
\label{algo:until t}
{\small
\begin{algorithmic}
\REQUIRE {$s, \tau, \gamma_1,\dots,\gamma_{\tau}$ and $t$}
\STATE{$i \leftarrow 0$ and $T_0\leftarrow 0$}
\REPEAT
\STATE{$i\leftarrow i+1$}
  \STATE{run Algorithm~\ref{algo:splitting2} to obtain the list $\c Y_{i}$ and its cardinality $M_{i}$}
	\STATE{$T_{i}\leftarrow T_{i-1}+M_{i}$}
\UNTIL{$T_{i} > t$}
\RETURN{$N(t)\leftarrow i$ and the empirical distribution $\widehat{\bb Q}_{N(t)}$
  of the set of states  $\c Y_\cup:=\c Y_1\cup\cdots\cup\c Y_{N(t)}$}
\end{algorithmic}
}
\end{algorithm}

Note that $\widehat{\bb Q}_{n}$ (or $\widehat{\bb Q}_{N(t)}$) is a random distribution; it is the 
distribution \emph{conditional} on $\c Y_\cup$.
The \emph{unconditional} distribution of a state obtained by generating $\c Y_\cup$
and then selecting one state randomly from $\c Y_\cup$ is also of interest:
this is the (a priori) distribution of a state sampled from $\widehat{\bb Q}_{n}$ (or $\widehat{\bb Q}_{N(t)}$),
but before we run the GS algorithm to construct $\c Y_\cup$.
We will denote these two unconditional distributions  by 
\[
  \bb Q_n(A) := \bb E [\widehat{\bb Q}_n(A)]\quad \textrm{(for Algorithm~\ref{algo:n iid})}
\]
and 
\[
  \overline{\bb Q}_t(A) := \bb E [\widehat{\bb Q}_{N(t)}(A)]
	\quad \textrm{(for Algorithm~\ref{algo:until t})}
\]
for all $A \in\c A$, where the expectation is with respect to the realization of $\c Y_\cup$. We saw earlier that 
\[
  \frac{\EE[\mathring{H}(A)]}{\EE[\mathring{M}]} = \bb P(\v Y \in A \mid \v Y \in B).
\]
Now let $H(A) = |\c Y \cap A|$ be the number of states returned by Algorithm~\ref{algo:splitting2}
that belong to $A$.   We have 
$\EE[{H}(A)] = \EE[\mathring{H}(A) \gvn \mathring{M} > 0] = \EE[\mathring{H}(A)] / \PP(\mathring{M} > 0)$.
Likewise, $\EE[M] = \EE[\mathring{M}] / \PP(\mathring{M} > 0)$.  Therefore,
\[
  \frac{\EE[{H}(A)]}{\EE[{M}]} 
	= \frac{\EE[\mathring{H}(A)]}{\EE[\mathring{M}]} 
	= \bb P(\v Y \in A \mid \v Y \in B).
\]
We also know that $\widehat{\bb Q}_n(A)$ and $\widehat{\bb Q}_{N(t)}(A)$
converge with probability one to $\EE[{H}(A)]/\EE[{M}]$ when $n\to\infty$ 
and when $t\to\infty$, respectively, from the strong law of large numbers applied to the 
numerator and the denominator.
Thus, they converge almost surely to the desired conditional probability 
$\bb Q(A) := \bb P(\v Y \in A \mid \v Y \in B)$.

\section{Convergence Analysis}
\label{sec:convergence}

We now analyze the convergence of the empirical distribution of the retained states,
$\widehat{\bb Q}_{n}$ (or $\widehat{\bb Q}_{N(t)}$), 
as well as its expected (i.e., unconditional on $\c Y_\cup$) version $\bb Q_n$ (or $\overline{\bb Q}_t$),
to the true conditional distribution $\QQ$.
The aim is to obtain non-asymptotic or risk bounds on the distance between $\QQ$ and the empirical distribution,
and its expected (unconditional) version. For a given class $\c A$ of measurable sets, we consider the three error criteria:
\begin{enumerate}
 \item The TV error between the expected (unconditional) distribution $\bb Q_n$ and $\bb Q$, that is:
\begin{equation*}
\sup_{A\in\c A} \left| \bb Q_n(A) - \bb Q(A)\right|\;.
\end{equation*}
This error measures the size of  the ``bias'' of  $\widehat{\bb Q}_n$ as an estimator of the true $\bb Q$. 
 \item The worst-case mean absolute error  of the conditional distribution $\widehat{\bb Q}_{n}$, defined as:
\[
\sup_{A\in\c A} \bb E\left| \widehat{\bb Q}_n(A) - \bb Q(A)\right|\;.
\]
\item The (random) TV error, $
 \sup_{A\in\c A}\left| \widehat{\bb Q}_n(A) - \bb Q(A)\right|
$, of the conditional distribution $\widehat{\bb Q}_{n}$, and its expected value:
\[
\bb E\sup_{A\in\c A}\left| \widehat{\bb Q}_n(A) - \bb Q(A)\right|\;.
\]
\end{enumerate}
By permuting the positions of the expectation, absolute value function,  and the supremum  ($\sup|\bb E[\cdot]|\rightarrow \sup\bb E|\cdot|\rightarrow \bb E\sup|\cdot|$),  we find that the three error criteria dominate each other as follows: 
\[
\overbrace{\bb E\sup_{A\subseteq \c A} |\widehat{\bb Q}_n(A)-\bb Q(A)|}^{\textrm{expected TV error}} ~\geq~ 
\sup_{A\subseteq \c A} \overbrace{\bb E|\widehat{\bb Q}_n(A)-\bb Q(A)|}^{\textrm{mean absolute error}} ~\geq~ 
\overbrace{\sup_{A\subseteq \c A} |\bb E[\widehat{\bb Q}_n(A)]-\bb Q(A)|}^{\textrm{TV error}}\;.
\]
In other words, the expected TV error is the  \emph{most stringent} 
of these three errors. In fact,  the (expected) TV error of the empirical distribution is so stringent that it does not converge, unless the class of sets  $\c A$ is  restricted. To ensure convergence, in Section~\ref{sec:random TV} we will take  $\c A$ to be  a restricted class of subsets. 
In contrast, the TV and mean absolute errors do not require any restrictions on the class $\c A$ and for these error criteria we simply take $\c A $ to be the class of all Borel subsets of $\bb R^d$.

\subsection{Convergence of Total Variation and  Mean Absolute Errors}
Let $m := \bb E[M]$ and $\var(M)$ denote the expectation and variance of $M$, which is the output of either Algorithm~\ref{algo:n iid}, or Algorithm~\ref{algo:until t}.
In this section, we state theorems giving non-asymptotic  bounds on the TV error and  the (worst-case) mean absolute error. The proofs of the following results are in the appendix.

\begin{theorem}[Sampling via $n$ iid runs of GS]
\label{thm:n iid}
The TV error is  bounded as 
\[
 \sup_{A}|\bb Q_n(A)-\bb Q(A)|\leq {c_1}{n^{-1}}
\]
where 
$
c_1 := \left({\var(M)+\sqrt{\var(M)\bb E[M^2]}}\right) {m^{-2}}.
$
The worst-case mean absolute error is bounded as
\[
\sup_A\bb E|\widehat{\bb Q}_n(A)-\bb Q(A)| \leq 
  {\tilde c_1(n)}{n^{-1/2}}
\]
where 
$
\tilde c_1(n):= \left({\sqrt{\bb E  M^2}}+ {\sqrt{3\bb E M^4 /n}}\right) m^{-1}
$
is bounded uniformly in $n$.
\end{theorem}

The terms $c_1$ and $\tilde c_1(n)$ in these bounds can be estimated from the simulation output.

\begin{theorem}[Sampling until GS returns $t$ states]
\label{thm:until t nonasymp}
In this case, the TV error is bounded as
\[
 \sup_{A}|\overline{\bb Q}_t(A)-\bb Q(A)| \leq {c_2(t)}{(t/m)^{-3/2}},
\]
where 
$
 c_2(t):= \sqrt{(4/3)\bb E [M^3] \bb E[M^2] \left(m+\bb E[M^2]/t\right)} \;m^{-3}
$
is bounded uniformly in $t$.
The worst-case mean absolute error is bounded as
\[
\sup_A\bb E|\widehat{\bb Q}_{N(t)}(A)-\bb Q(A)|\leq \tilde c_2(t)(t/m)^{-1/2},
\]
where $\tilde c_2(t):= ({\bb E M^2})^{1/2} m^{-1} + ({\bb EM^2}) m^{-3/2} t^{-1/2}$ 
is also uniformly bounded in $t$.
\end{theorem}

Again, the terms $c_2(t)$ and $\tilde c_2(t)$ can be estimated easily by simulation: 
it suffices to estimate ${\bb E M^2}$ and ${\bb E M^3}$ by their empirical versions.
The constant $m$ in $(t/m)^{-3/2}$ could be absorbed into $c_2(t)$, 
but we choose not to do this, because we want to be able to compare $\overline{\bb Q}_t$ and 
$\bb Q_n$ on a common scale, where $n$ (the simulation effort of Algorithm~\ref{algo:n iid}) and $t/m$ (the average simulation effort of Algorithm~\ref{algo:until t} for large $t$) are the same. 
The key point to notice is that we get a better rate for the bound for 
$\overline{\bb Q}_t$ than for $\bb Q_n$.

In the next result, we obtain an improved convergence rate of $\c O(1/t^2)$,
but at the price of introducing in the bound an $\c O(\exp(-\omega t))$ term 
(for some $\omega > 0$) which is hard to estimate.
This term converges exponentially fast in $t$, so it is asymptotically negligible 
when $t\to\infty$, but it is not necessarily negligible for a given (finite) $t$.
So we have an asymptotically better bound that we cannot easily estimate.
In practical settings, we may prefer the $\c O((t/m)^{-3/2})$ bound from 
Theorem~\ref{thm:until t nonasymp} that we can more easily estimate to the 
$\c O((t/m)^{-2})$ bound that we cannot completely estimate.

\begin{theorem}[Sampling until GS returns $t$ states; asymptotic version]
\label{thm:until t}
We have
\[
 \sup_{A}|\overline{\bb Q}_t(A)-\bb Q(A)|\leq  c_3 (t/m)^{-2} + \c O(\exp(-\omega t)),
\]
where $\omega>0$ is a (typically unknown) constant and 
\[
 c_3:=\frac{\bb E[M^2|M-1-2r|]}{2m^3}
\] 
with $r:=(\bb EM^2+m)/(2m)$.
\end{theorem}

This result does not include a statement about the mean absolute error, because   the bounds of the mean absolute errors in Theorems~\ref{thm:n iid} and \ref{thm:until t nonasymp} already converge at the optimal asymptotic rate,  and thus cannot be improved.

\subsection{Convergence of the Empirical Conditional Distribution $\widehat{\bb Q}_n$}
\label{sec:random TV}

We now examine the convergence of the TV error between the empirical distribution 
$\widehat{\bb Q}_n$ and $\bb Q$ when $n\to\infty$.
This distribution is random, and any realization is discrete with finite support,
so obviously it cannot converge to $\bb Q$ in TV with $\c A$ taken as all the Borel sets,
because by taking $A$ as the finite set $\c Y_\cup$,
we get $\widehat{\bb Q}_n(A) = 1$ for any $n$, but $\bb Q(A) = 0$ (assuming that $\QQ$ has a density).
Thus, as mentioned previously, we necessarily have to restrict the class $\c A$. We start by giving conditions for TV convergence with probability 1 under the following restrictions on the class $\c A$.

\begin{assumption}
\label{ass:TVwp1}
Suppose that one of the following two conditions holds:
\begin{enumerate}
\item $\c A$ is a class with finite Vapnik-Chervonenkis (VC) dimension, or
\item $\c A$ is the class of all convex sets in $\bb R^d$, and the 
 transition kernel  in Algorithm~\ref{algo:splitting}  has a probability density $\kappa_l(\v y\mid \v x)$.
\end{enumerate}
\end{assumption}

\begin{theorem}[Almost-Sure TV Convergence]
\label{theorem:a.s.} 
Under Assumption~\ref{ass:TVwp1}, we have almost sure TV convergence:
\[
 \sup_{A\subseteq \c A}|\widehat{\bb Q}_n(A)-\bb Q(A)|\longrightarrow 0 \quad 
   \textrm{ with probability 1 when } n\to\infty.
\]
\end{theorem}

The notion of VC dimension is discussed for example by \cite{vapnik2013nature}.
Roughly speaking, it measures the flexibility of a class of subsets to correctly classify data defined over $\RR^d$, and in our context it measures the complexity of the class of sets $\c A$. Sets with higher VC dimension are more complex.

Note that the class of convex sets has an infinite VC dimension,
which is why the second option in Assumption~\ref{ass:TVwp1} requires the extra regularity condition
on the transition kernel.  This condition will be satisfied  if $\kappa_l$ 
is the transition kernel of a Gibbs sampler, but will not be satisfied for the kernel of a 
Metropolis-Hastings sampler \citep[Equation 6.3, Page 226]{kroese2013handbook}.
Note that the condition does not require that we have a closed form (simple) formula for the transition density 
$\kappa_l(\v y\mid \v x)$.  It only requires that it exists.

Our next result (proof in Appendix~\ref{append:theorem 1}) provides bounds on the expected TV error of the empirical distribution, where $\c A$ is a class of sets with a finite VC dimension.
\begin{theorem}[Bound on Expected TV for Empirical Distribution]
\label{theorem:simple}
Suppose the class $\c A$ has finite VC dimension $v$. 
Then, the expected TV error made by using the empirical distribution $\widehat{\bb Q}_n$ 
as an approximation of $\bb Q$ is bounded as follows:
\[
 \bb E\sup_{A\in \c A}|\widehat{\bb Q}_n(A)-\bb Q(A)|
 \leq \frac{\sqrt{\var(M)}}{m\sqrt{n}}+\frac{2\sqrt{v\,\ln(2n)\,\bb E [M^2\ln M]}}{m\sqrt{n}}\;\psi_1(v,n),
\]
where
\[
 \psi_1 = \psi_1(v,n) := \sqrt{\frac{(\ln(2)+v+v\ln (2n/v))\bb E[ M^2]}{v\ln(2n)\bb E [M^2\ln M]}
               + \frac{1}{\ln(2n)}} < \infty
\]
is bounded uniformly in $(v,n,\tau)$.
\end{theorem}

As an example, let $[\v a, \v b] = \{\v y\in\RR^d : \v a\leq\v y\leq \v b\}$ 
represent a rectangle in $\bb R^d$, 
and suppose $\c A$ is the class of all rectangles in $ \bb R^d$.  Then $v=2d$ \citep{sauer1972density}.
If $\v a=-\v\infty$, that is, $\c A$ as the class of one-sided intervals of the form
$[-\v\infty, \v b]$, then $v=d+1$.
In this case, the previous theorem can provide a bound on the expected value of the Kolmogorov-Smirnov statistic:
\begin{equation}
\label{KS statistic}
\mathrm{ks}(n):=\bb E\left[\;\sup_{\v x\in \bb R^d} 
     \left|\widehat{\bb Q}_n(\v X\leq \v x)-\bb Q(\v X\leq \v x)\right|\;\right].
\end{equation}
We will use this type of error bound in Section~\ref{sec:convergence assessment} when we assess the quality of 
our approximate sampling from a Bayesian posterior.

Using the metric entropy of the class $\c A$, it is also possible to obtain a  bound without 
the logarithmic growth term $\ln (n)\ln(M)$  in Theorem~\ref{theorem:simple}, 
and to get an expected TV bound 
that depends solely on the relative second moment of $M$.

\begin{theorem}[Second Bound on Expected TV for Empirical Distribution]
\label{theorem:hard}
Let $\tau$ be the number of levels in Algorithm~\ref{algo:splitting} with splitting factor $s$ and 
suppose that $\c A$ has VC dimension $v<\infty$.
Then the empirical distribution $\widehat{\bb Q}_n$ satisfies:
\[
  \bb E\sup_{ A\in \c A}|\widehat{\bb Q}_n(A)-\bb Q(A)|  \leq
  \frac{\sqrt{\var(M)}}{m\sqrt{n}}+\frac{(s+1)4\sqrt{v\;\bb E[M^2]}}{m\sqrt{n}}\psi_2(\tau,v,n,s),
\]
where
\[
 \psi_2 = \psi_2(\tau,v,n,s) := \sum_{k=1}^{\lceil\tau+\log_s\sqrt{n}\rceil} 
  \frac{1}{s^{k}} \left(\frac{\ln 2}{2nv }+\frac{1+\ln(v+1)}{v}+1+\ln(2s^{2k})\right)^{1/2}  ~<~ \infty
\]
is bounded uniformly in $(\tau,v,n,s)$.
\end{theorem}

Unfortunately, as we shall see in Section~\ref{sec:convergence assessment}, 
the constant $\psi_2$  in this bound is much larger than $\psi_1$ in Theorem~\ref{theorem:simple}. 
As a result, $n$ has to be impractically large for the above bound to beat the  
simpler bound in Theorem~\ref{theorem:simple}. 
Nevertheless, the result is still of theoretical interest as it shows that the rate of convergence 
in expectation of the TV distance can be improved from 
$\c O(\ln(n)/\sqrt{n})$ to the canonical rate of $\c O(1/\sqrt{n})$. 
In addition, the term $\bb E[M^2\ln M]/m^2$ in Theorem~\ref{theorem:simple} does not appear 
in Theorem~\ref{theorem:hard}.

\begin{remark}[Simplifications due to Existence of a Density]
 If the transition density $\kappa_l(\v x|\v y)$
is available in closed form and easily evaluated, we can do much
better by dropping the restrictions that the class $\c A$ has a finite VC dimension. Instead, if  $\kappa(\v x|\v y)\equiv \kappa_{\tau}(\v x|\v y)$ is a transition density  with stationary pdf $q$,
then we can define the  empirical density:
\[
\hat q_n(\v x)\idef \frac{1}{T_n} \sum_{\v Y\in \c Y_\cup} \kappa(\v x|\v Y),
\]
so that we can use Sheff\'e's  identity \citep[Theorem 5.1]{devroyecombinatorial} to simplify the uniform deviation over the class $\c B$ of Borel measurable sets:
\[
2\sup_{A\in \c B}\left|
\int_A \hat q_n(\v x)\di \v x-\bb Q(A)\right|=\int_{\bb R^d} |\hat q_n(\v x)-q(\v x)|\di \v x
\]
Therefore, the bound on the expected TV distance simplifies as follows:
\[
\begin{split}
 2\bb E\left[\sup_{A\in \c B}\left|
 \int_A \hat q_n(\v x)\di\v x-\bb Q(A)\right|\right]
&\leq  
\bb E\int_{\bb R^d}\left|\frac{1}{n \bar M_n} \sum_{\v Y\in \c Y_\cup} \kappa(\v x|\v Y)
  - \frac{1}{n m} \sum_{\v Y\in \c Y_\cup} \kappa(\v x|\v Y) \right|\di\v x\\              
&\qquad+\bb E\int_{\bb R^d}\left|\frac{1}{n m} \sum_{\v Y\in \c Y_\cup} \kappa(\v x|\v Y) -q(\v x)\right|\di\v x\\
&\leq  \frac{1}{m\sqrt{n}}\Big(\sqrt{\var(M)} + \textstyle \sqrt{\int_{\bb R^d}\var(\kappa(\v x|\v Y))\di \v x}\Big)\;.
\end{split}
\]
Thus, provided   the integrated variance $\int\var(\kappa(\v x|\v Y))\di \v x$ can be estimated easily, this bound can be used as a simpler alternative to Theorem~\ref{theorem:simple}. We do not pursue 
this possibility further in this article.
\end{remark}

\section{Numerical Example: Bayesian Lasso}
\label{sec:numerics}
\label{sec:example}

In this section we consider an application of the splitting sampler in Algorithm~\ref{algo:splitting2} 
to the problem of posterior simulation in Bayesian inference. 
We estimate the bounds in Theorems 1 to 6 in order to assess the convergence of Algorithms~\ref{algo:n iid} and \ref{algo:until t}. This convergence assessment can be used  to either assess whether any Bayesian credible intervals are reliably estimated from the simulation output, or to rank the performance of implementations that use different Markov chain kernels $\kappa_l$ (the Markov chain that yields the smallest TV error will be the preferred one).

\subsection{Approximate Posterior  Simulations via Splitting}

One of the simplest and most widely used linear regression models 
for data $\v y=(y_1,\ldots,y_{n'})^\top$ is the \emph{Bayesian Lasso} \citep{park2008bayesian},
in which the  point-estimator of the 
regression coefficient $\v\beta = (\beta_1,\dots,\beta_d)^\top \in \bb R^d$ 
is defined as the minimizer of the constrained least squares problem:
\[
 \min_{\v\beta}\|\v y -\m X\v \beta\|^2_2,\qquad \textrm{subject to }
\|\v\beta\|_1\leq \gamma,
\]
where: (1) $\m X$ is a matrix with $d$ columns (predictors); (2) the   term 
$\|\v\beta\|_1=|\beta_1|+\cdots+|\beta_d|$ is the \emph{least absolute shrinkage and selection operator} (Lasso); and  (3) $\gamma\geq 0$ is the  Lasso regularization parameter. 
In a Bayesian linear regression,  one wishes to estimate the
posterior distribution of the parameters $(\v\beta,\sigma^2)$, that is, the distribution of $(\v\beta,\sigma^2)$ conditional on the data $\v y$ and the constraint $\|\v\beta\|_1\leq \gamma$. Since this posterior distribution is intractable, one typically approximates it by sampling random pairs $(\v\beta,\sigma)$ 
from the posterior pdf:
\begin{equation}
\begin{split}
\label{eq:bayes-posterior}
 q(\v\beta,\sigma|\v y, \gamma)
   &~\varpropto~ \phi(\v y-\m X\v\beta;\sigma^2\m I)\;\sigma^{-2}\;
     \bb I\{\|\v \beta\|_1\leq\gamma\} \times d!/{(2\gamma)^d}\;,
\end{split}
\end{equation}
where: a) $\phi(\v x;\Sigma)$ denotes the multivariate normal pdf 
with mean zero and covariance matrix $\Sigma$ evaluated at $\v x$; 
b) the factor $\sigma^{-2}$ results from using an uninformative prior for the scale $\sigma$, and 
c) ${\bb I\{\|\v \beta\|_1\leq\gamma\}} \times d!/{(2\gamma)^d}$ 
is the  prior of $\v\beta$, uniform over the feasible set.
Note that, unlike the  more
 common Laplace prior used in the Bayesian Lasso \citep{park2008bayesian}, here the prior enforces the constraint on $\v\beta$ directly.

To sample a new state $(\v\beta_k,\sigma_k)$ during the course of splitting, we need to simulate from a  transition density $\kappa_l((\sigma_{k},\v\beta_{k})\gvn (\sigma_{k-1},\v\beta_{k-1}))$, which is stationary with respect to the density \eqref{eq:density-ft}.  We simulate a move from $(\sigma_{k-1},\v\beta_{k-1})$ to $(\sigma_{k},\v\beta_{k})$ as follows.  Given $\v\beta_{k-1}$, we sample
\[
  (1/\sigma^2_{k})\ \sim\ \mathsf{Gamma}((n'+1)/2,\|\v y-\m X\v\beta_{k-1}\|^2_2/2),
\] 
which is the gamma distribution with mean  $(n'+1)/\|\v y-\m X\v\beta_{k-1}\|^2_2$
and shape parameter $(n'+1)/2$.
Given $(\sigma_{k},\v\beta_{k-1})$, we simulate $\v\beta_k$ via a ``hit-and-run"
Gibbs sampler \cite[Page 240]{kroese2013handbook}. 
In other words, the new state is $\v\beta_{k}=\v\beta_{k-1}+\lambda\v d$, where $\v d$ is a point uniformly distributed on the surface of the $d$-dimensional  unit hyper-sphere, and
the scalar $\lambda$ is simulated according to:
\[
 (\lambda\gvn\v d,\sigma_k,\v\beta_{k-1}) 
 ~\sim~ \varphi(\lambda\gvn\v d,\sigma_k,\v\beta_{k-1}) 
 ~:=~  \frac{q(\v\beta_{k-1}+\lambda\v d,\sigma_k\gvn \v y, \gamma)}{
       \int q(\v\beta_{k}+\xi\v d,\sigma_k\gvn \v y, \gamma)\di\xi}\;.
\] 
 The conditional pdf  $\varphi(\lambda\gvn\v d,\sigma,\v\beta_{k-1})$ 
is a univariate truncated normal, which can be simulated easily \citep{valuetools2016}.


As a concrete illustration we use the  ``diabetes  dataset" 
 \citep{park2008bayesian}, consisting of $n'=442$  patients. For each patient, we have a 
record of $d=10$ predictor variables (age, sex, body 
mass index, and 7 blood serum measurements, so that $\m X$ is a matrix
of size  $442\times 10$), and a response variable, which 
measures the severity of nascent diabetes. We fix 
$\gamma=1200$, which  corresponds to the Lasso regularization parameter value used 
by \cite{park2008bayesian}.

\begin{figure}[htb]
\begin{center}
\includegraphics[scale=.8]{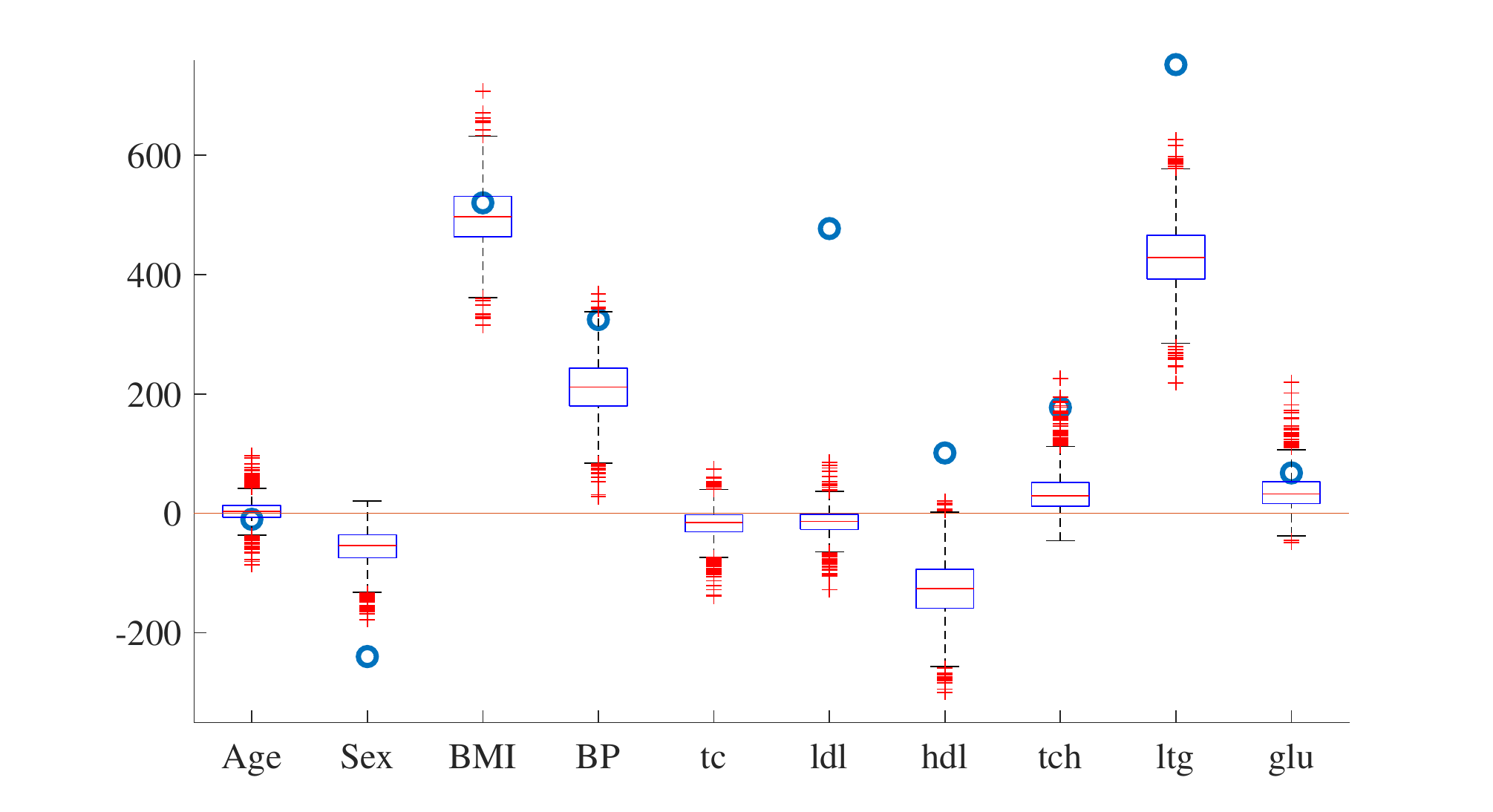}
\end{center}
\caption{Empirical marginal distributions of the ten coefficients $\beta_j$ corresponding to the ten
 predictors, sampled approximately from \eqref{eq:bayes-posterior}. 
 For comparison, the unconstrained (ordinary) least squares solution for each $\beta_j$ is displayed as a circle.}
\label{fig:million16h}
\end{figure}

To simulate from the Bayesian posterior \eqref{eq:bayes-posterior} we ran Algorithm~\ref{algo:splitting2}
with splitting factor $s=100$  and $n=10^4$ using the following $\tau=4$ levels:
$
  (\gamma_1,\gamma_2,\gamma_3,\gamma_4)  = (1907,         1368,       1230,         1200)
$
to obtain the multiset $\c Y_\cup$.
The first three levels were chosen so that $\rho_j\approx 0.01$ for $j=1,\ldots,3$. 
The values for $\tau,\gamma_1,\ldots,\gamma_\tau$ were selected by running the adaptive pilot 
algorithm in \citep{botev2012dependent}[Algorithm 4]. The marginal  empirical distribution  of each coefficient $\beta_j$
is illustrated in Figure~\ref{fig:million16h} as a boxplot. 

\subsection{Convergence Assessment via Theoretical Bounds}
\label{sec:convergence assessment}

\begin{figure}[htb]
\begin{center}
\includegraphics[width=1\linewidth]{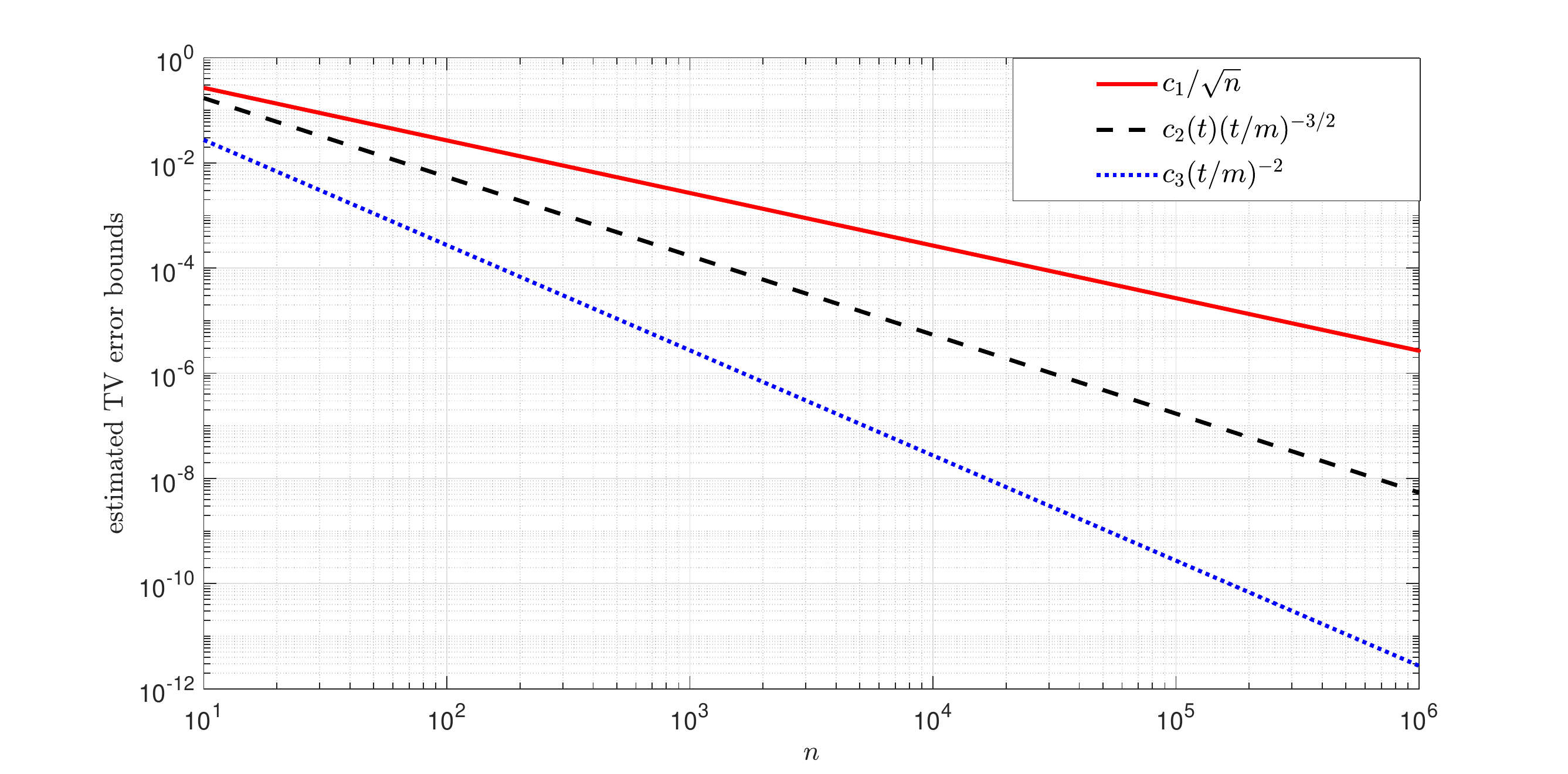}
\end{center}
\caption{Comparison of three  bounds on the TV error, $c_1/\sqrt{n},\;c_2(t)\times(t/m)^{-3/2}$, and $c_3\times(t/m)^{-2}$, where $t=n\times m$.}
\label{fig:TVbounds1}
\end{figure}

Using the output of Algorithm~\ref{algo:splitting2} from the previous section we calculated point estimates of  the unknown  terms, $c_1,\tilde c_1(n),c_2(t),\tilde c_2(t),c_3,\psi_1,\psi_2$, appearing in  Theorems 1 through 6. Note that all the unknown terms depend on moments of $M$. For example, some of the point-estimates of the moments of $M$ are $(\bb EM,\bb EM^2)\approx(5.9 , 71)$. 
 Figure~\ref{fig:TVbounds1} shows the estimates 
of $c_1/n$, $c_2(t)\, (t/m)^{-3/2}$, and $c_3\, (t/m)^{-2}$, 
which bound  the TV error (see Theorems 1 to 3),
 on a common scale with $t= n \times 5.9$ (since $m=\bb E[M]\approx 5.9$).

There is one major take-home message from  Figure~\ref{fig:TVbounds1}, namely, that Algorithm~\ref{algo:until t} (sampling to exceed $t$ states)  simulates more closely (in terms of TV error) from the target distribution $\bb Q$ than Algorithm~\ref{algo:n iid} ($n$ iid non-empty replications). Of course, the downside of using Algorithm~\ref{algo:until t} is that the number of trials, $N(t)$, is random (with expectation  $t/m$ for large $t$).

In addition, reading off from Figure~\ref{fig:TVbounds1} we can see that if we run Algorithm~\ref{algo:until t} with  $t>5.9\times 10^3$, then the TV error between $\overline{\bb Q}_t$ and $\bb Q$ is estimated as less than $10^{-3}$ using the non-asymptotic bound
$c_2(t)(t/m)^{-3/2}$ and as less than $10^{-5}$ using the asymptotic bound $c_3(t/m)^{-2}$ (it is asymptotic, because we ignored the asymptotically negligible  $\c O(\exp(-\omega t))$ term in Theorem~\ref{thm:until t}). 

\begin{figure}[htb]
\begin{center}
\includegraphics[width=.5\linewidth]{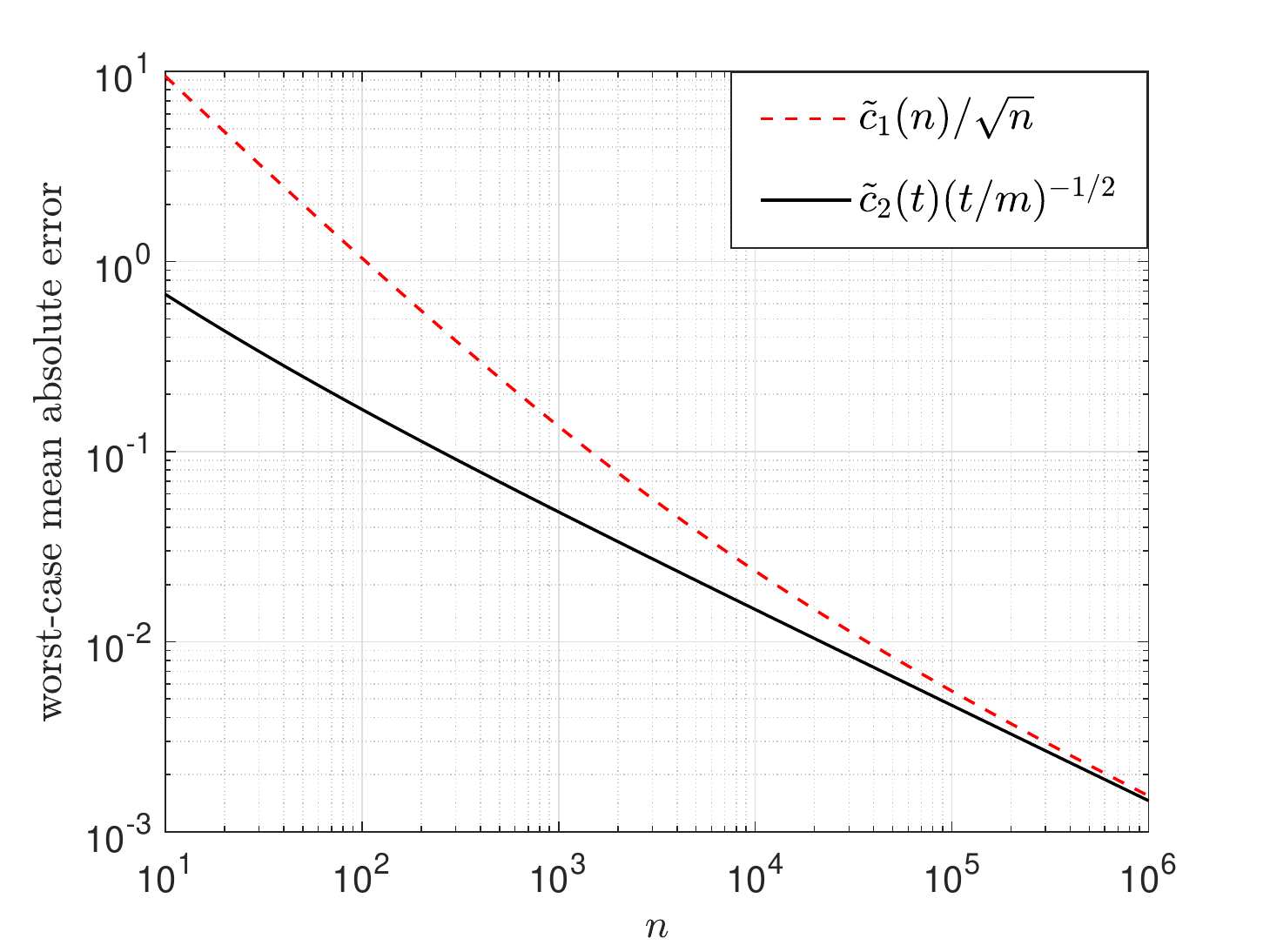}~\includegraphics[width=.5\linewidth]{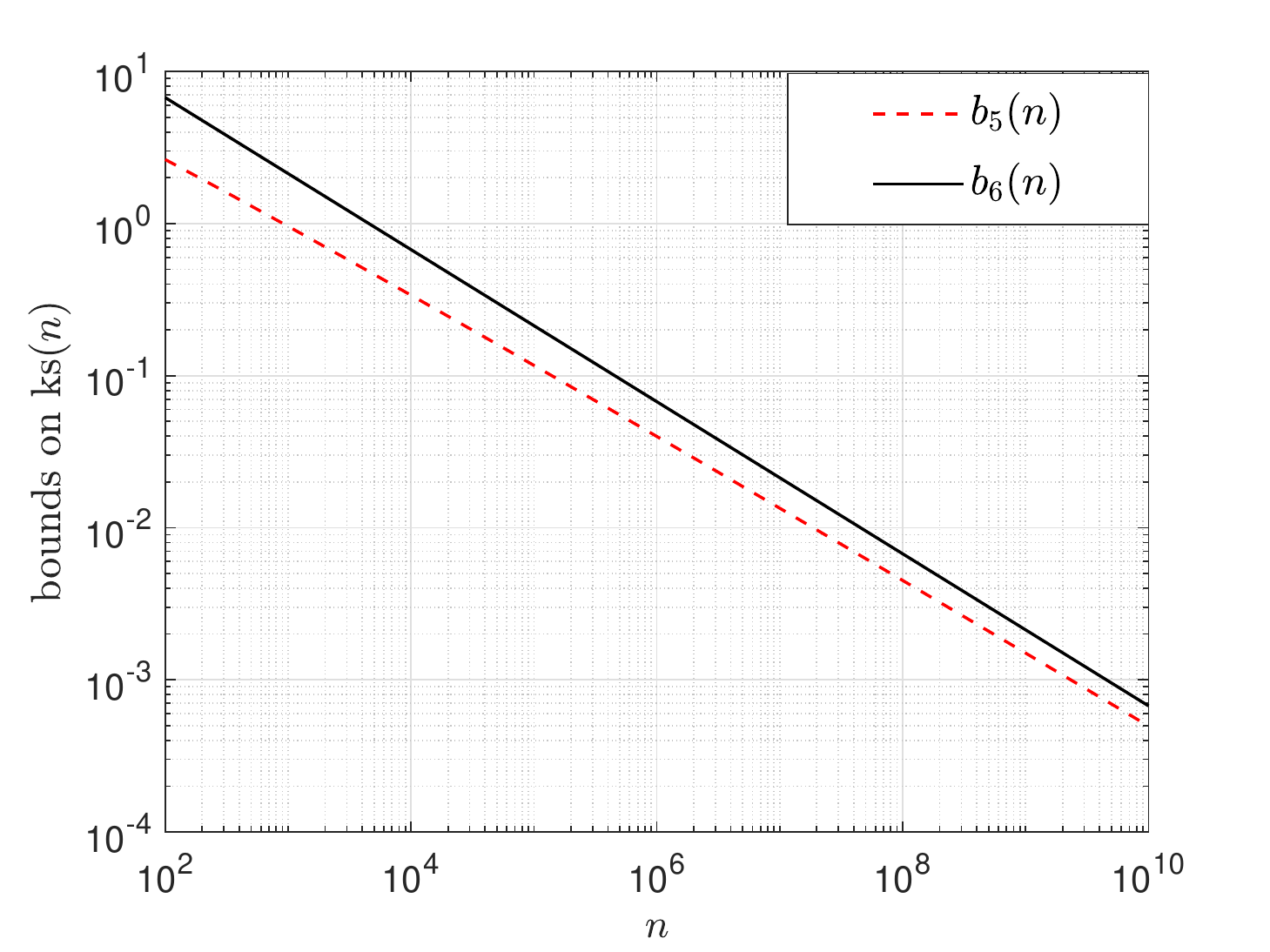}
\end{center}
\caption{Left: estimates of the worst-case mean absolute error; Right: estimates of the expected TV error.}
\label{fig:risk bnds}
\end{figure}

As for the mean absolute error, the left pane of  Figure~\ref{fig:risk bnds} shows the estimated bounds 
$\tilde c_1(n) {n}^{-1/2}$ and $\tilde c_2(t)(t/m)^{-1/2}$ given in Theorems~\ref{thm:n iid} and \ref{thm:until t nonasymp}, respectively, using $t=n\times m$. 

 It is clear that the bound $\tilde c_2(t)(t/m)^{-1/2}$ is always  smaller. Note that both bounds are asymptotically equivalent to first order ---  as $n$ becomes larger, the two bounds converge to each other. Based on the mean absolute error, in this example we again conclude that Algorithm~\ref{algo:until t}
(sample more than $t=m\times n$ states) is a better performing sampler  than Algorithm~\ref{algo:n iid} ($n$ iid non-empty runs).

Next, we apply  the results of Theorems~\ref{theorem:simple} and \ref{theorem:hard} to bound the expectation of the
 Kolmogorov-Smirnov statistic, $\mathrm{ks}(n)$, given in \eqref{KS statistic}. Let $b_5(n)$ and $b_6(n)$ be the upper bounds on \eqref{KS statistic} derived in  Theorems~\ref{theorem:simple} and \ref{theorem:hard}, respectively (here $v=d+1=11$). The right pane of Figure~\ref{fig:risk bnds} shows the estimated bounds on the value of $\mathrm{ks}(n)$. There are a number of observations to be made. 

First, we can see that for the range of the plot, $b_5(n)=\c O(\ln(n)/\sqrt{n})$ yields a better risk bound than $b_6(n)=\c O(1/\sqrt{n})$ (despite the superior convergence rate of $b_6$). This is because, as mentioned previously, the constant $\psi_2$ in Theorem~\ref{theorem:hard} is much larger than $\psi_1$ in Theorem~\ref{theorem:simple}. In fact, the cross-over  for which ultimately $b_6(n)<b_5(n)$ happens for $n>10^{19}$ (not shown on Figure~\ref{fig:risk bnds}).
 
Second, from the right pane of Figure~\ref{fig:risk bnds} we can see that the expectation of the
 Kolmogorov-Smirnov statistic is indeed the most stringent  error criteria, because we need a very large $n$ to guarantee an acceptably small error (at least $n>10^7$ to make $b_5(n)$ smaller than $10^{-2}$).

Third, we observe that since the transition kernel, $\kappa_l$, has a density (it is the transition pdf of a  Gibbs Markov chain), Theorem~\ref{theorem:a.s.} ensures the almost sure convergence 
of the empirical TV uniformly over the class $\c A$ of all convex subsets, that is,
$
 \sup_{A\in \c A}|\widehat{\bb Q}_n(A)-\bb Q(A)|\longrightarrow 0
$ with probability one.

Finally, we note that our convergence results do not theoretically quantify the  speed of convergence of the Markov chains, induced by the kernels $\kappa_l$. This dynamics  is captured by  the moments of $M$, which we estimate empirically, but not theoretically. To analyze theoretically the growth of  the moments of $M$ will require  an analysis of the speed of convergence of all Markov chains used in Algorithm~\ref{algo:splitting}.

\subsection{Comparison with Sequential Monte Carlo for Rare Event Estimation}

In the Bayesian context,  the rare-event probability 
$\ell(\gamma)=\bb P(\|\v\beta\|_1\leq\gamma)$  is the normalizing constant of the  posterior \eqref{eq:bayes-posterior}, also called the \emph{model evidence} or \emph{marginal likelihood}, which is of importance in model selection and inference. 

From equation \eqref{hatell} above, 
we can see that an estimator of $\ell$ using  $n=10^4$ independent runs of Algorithm~\ref{algo:splitting} is 
$\hat\ell:=(\mathring{M}_1+\cdots+\mathring M_n)/(n s^{\tau-1})$ with relative error
$\sqrt{\var(\mathring M)}\big/(\hat\ell\sqrt{n})$. We obtained the estimate of $\hat\ell=2.4\times 10^{-8}$ with estimated relative error of $3.6\%$.

For completeness, and as a  benchmark  to our results, we  compared the performance of Algorithm~\ref{algo:splitting}  with the popular  sequential Monte Carlo (SMC) method for rare-event estimation of  
\cite{cerou2012sequential}, as described on top  of page 798, column 1.
For the SMC we used the same intermediate thresholds $
  (\gamma_1,\gamma_2,\gamma_3,\gamma_4)  = (1907,         1368,       1230,         1200)
$ (in the notation on page 798, we have $A_k:=\{\|\v\beta\|_1<\gamma_{k}\}$) and a total simulation effort of $6\times 10^6$ particles to estimate 
 $\ell$. This is roughly twice the average simulation effort for $n$ runs of Algorithm~\ref{algo:splitting}, which is approximately $n\times\sum_{k=1}^\tau\frac{1}{\rho_k}\approx 3.4\times 10^6$.  Despite this, the relative error of the SMC estimator of $\ell$ was estimated as $12\%$, or about three times larger than the relative error of $\hat\ell$.

The  observation that the GS algorithm can, under certain conditions, 
perform better than sequential Monte Carlo methods is known and is already explained in
\cite{botev2012efficient}.  Briefly, the GS sampler is expected to outperform standard SMC methods when the Markov chain induced by $\kappa_l$ converges slowly to its stationary pdf \eqref{eq:density-ft}. Conversely, when the Markov chain at each level $l$ mixes fast (the particles follow the law of \eqref{eq:density-ft} almost exactly), then SMC methods  are to be preferred. As previously explained  \citep{botev2012efficient},  unlike standard SMC methods, the GS sampler does not have a bootstrap resampling step, which is advantageous when the transition kernel $\kappa_l$ fails to create enough ``diversity'' in the samples (bootstrap resampling reduces the diversity). This advantage, however, disappears if the  Markov chains at each level are mixing fast, and as a result using a fixed number of particles at each level \citep[Page 798]{cerou2012sequential} leads to superior accuracy compared to using a random number of particles (as in the GS Algorithm~\ref{algo:splitting}).

\section{Summary and Conclusions}
\label{sec:conclusion}

We  presented two different implementations  of the generalized splitting method that
can be used to simulate approximately from a conditional density in high dimensions. 
In the first implementation, we construct an empirical distribution 
$\widehat{\bb Q}_n$ from $n$ iid non-empty replications of the GS sampler (Algorithm~\ref{algo:splitting}). In the second implementation, we  construct an empirical distribution 
$\widehat{\bb Q}_{N(t)}$ by running Algorithm~\ref{algo:splitting2} until we have more than $t$ states in total. 
In both implementations,  $\widehat{\bb Q}_n$ and 
$\widehat{\bb Q}_{N(t)}$, and their respective expectations $\bb Q_n$ and $\overline{\bb Q}_t$, aim to  approximate the true distribution $\bb Q$. 

To assess the quality of the approximations we derived non-asymptotic bounds on three different error criteria: 
(1) the total variation  errors of $\bb Q_n$ and $\overline{\bb Q}_t$,   widely used in MCMC convergence analysis; (2) the mean absolute errors of $\widehat{\bb Q}_n$ and $\widehat{\bb Q}_{N(t)}$; 
and (3) the expected total variation error of $\widehat{\bb Q}_n$.

The main take-away messages are as follows. 
First, the GS sampler in Algorithm~\ref{algo:until t}, which samples until we have more than $t$ states in total, converges faster  than the GS sampler in Algorithm~\ref{algo:n iid}, which samples  $n$ iid  non-empty replications. 

Second,  the  proposed splitting samplers  provide a simple qualitative method for assessing whether  they are sampling accurately from the target distribution. Any  unknown constants and terms in the theoretical error estimates  depend only on moments of the number $M$ of particles, which can be readily estimated from the simulation output. This allows us to make qualitative statements such as ``choose $n>10^3$ to (approximately) obtain  a total variation error of less than $10^{-3}$'', or 
 to  rank  the performance of different implementations of the algorithms.

Finally, we have confirmed that, under certain conditions, generalized splitting can be more efficient than  sequential Monte Carlo in estimating rare-event probabilities.  This observation extends not just to estimation, but approximate sampling as well, because if an algorithm is not the most efficient in estimating a rare-event  probability, then it will also not be the most efficient algorithm to simulate conditional on the rare event.

\vskip 40pt
\begin{APPENDICES}

\section{Proof of the Theorems}
\label{sec:proofs}

We first recall the working notation.
Let $\c A$ be a class of measurable sets. 
For any  $ A\subseteq \c A$ and $i\ge 1$, let 
$M_i$ and $H_i(A)$ be the cardinalities of $\c Y_i$ and of $\c Y_{i} \cap  A$.
These are the realizations of $M$ and $H(A)$ for replication $i$ of Algorithm~\ref{algo:splitting2}.
Let $\bar M_n$ and $\bar H_n(A)$ be the respective averages of these $n$ realizations,
and let $m \idef \EGS[M]$, so that the target distribution is $\bb Q( A)=\EGS[H( A)]/m $.
For simplicity of notation, unless there is ambiguity, we henceforth drop the GS subscripts from $\EGS$.
When we draw an  $\v Y$ from $\widehat{\bb Q}_n$, 
it belongs to $A$ with probability $\bar H_n( A)/\bar M_n$ (since $\c Y_\cup$ is not empty,  $\bar M_n > 0$).
Note that $H(A) \leq M$ for all $A\in \c A$, and that $M_i$ and $H_i$ take their values in $\{1,...s^{\tau-1}\}$.

In particular,  in  Algorithm~\ref{algo:n iid} we obtain the independent sets,  $\c Y_{1}, \c Y_{2},\cdots,\c Y_{n}$, of  states $\v Y$.  We can (re)label all the states $\v Y$ such that:
 \[
\overbrace{\v Y_1,\ldots,\v Y_{T_1}}^{\c Y_1}, \ldots\ldots\ldots,
\overbrace{\v Y_{T_{n-1}+1},\ldots,\v Y_{T_n}}^{\c Y_n}\;.
\]
In this way,
  $\{\v Y_t, t=1,2,\ldots\}$ is
a discrete-time  regenerative process with regeneration times
$
0=T_0 < \ldots< T_n,
$
and tour lengths $M_i=T_{i}-T_{i-1}, \;j=1,2,\ldots,n$ with stationary measure $\bb Q(A)$.
With this notation we have that $N(t) = \min\{n : T_n > t\}$ in Algorithm~\ref{algo:until t}. Moreover,
if we define the number of renewals in $(0,t]$ as $\widetilde N(t) := N(t)-1 = \max\{n : T_n \leq t\}$ with $N(0)=0$, then $\{\widetilde N(t),\, t\ge 0\}$
 is a \emph{renewal process} \cite[Chapter 5]{asmussen2008applied}.

Since $N(t)=\widetilde N(t)+1$ is a stopping time with respect to the 
filtration generated by the sequence of iid random variables $\{M_i,\, i\ge 1\}$,
by the Wald identity we have 
$
\bb E [T_{N(t)}] = \bb E[N(t)]\bb E [M_i].
$
We define
$\widehat{\bb Q}_{n}(A)= \bar H_n(A)/{\bar M_n}$
and 
$
 \widehat{\bb Q}_{N(t)}(A) = \frac{1}{T_{N(t)}} \sum_{i=1}^{N(t)} H_i(A).
$
With $Z_i(A) := H_i(A)-M_i\bb Q(A)$, Wald's identity also gives
\begin{equation}
\label{Wald}
  \bb E\left[\sum_{i=1}^{N(t)}Z_i(A)\right] = \bb E[N(t)]\bb E [Z_i(A)] = 0.
\end{equation}
\begin{remark}[Elapsed-time process]
Note that the autocorrelation plot of the age (or current lifetime) process, 
 $E(t):=t-T_{\widetilde N(t)}$, may be used as a  graphical tool to diagnose the convergence of $\{\v Y_t, t=1,2,\ldots\}$  to its stationary distribution $\bb Q(A)$,  because \citep[Page 170, Proposition 1.3]{asmussen2008applied}:
\[
\sup_A|\bb P(\v Y_t\in A)-\bb Q(A)|\leq 2\sup_A|\bb P(E(t)\in A)-\bb P(E(\infty)\in A)|.
\]
In other words,
ensuring the convergence of the Markov process $\{E(t),t\geq 0\}$ to its stationary measure is sufficient to ensure the
convergence of $\{\v Y_t, t=1,2,\ldots\}$ to its stationary measure. 
\end{remark}

\subsection{Proof of Theorem~\ref{thm:n iid}}

First, we prove the bound on the TV error. Using the identity, \cite[Page 180]{sMEK82a}
\begin{equation}
\label{bias identity}
\frac{x}{y}-\frac{\alpha}{\beta}=\frac{x}{y}\frac{(y-\beta)^2}{\beta^2}+\frac{x}{\beta}-\frac{\alpha y}{\beta^2}-\frac{(x-\alpha)(y-\beta)}{\beta^2},
\end{equation}
with $\alpha=0$, we have that 
\[
\begin{split}
\bb E\left[\frac{\bar H_n(A)}{\bar M_n}\right]-\frac{h(A)}{m}&=\bb E\left[\frac{\bar Z_n(A)}{\bar M}\right]\\
&=\bb E\left[\frac{\bar Z_n(A)}{\bar M_n}\frac{(\bar M_n-m)^2}{m^2}\right]-\frac{\cov(\bar Z_n(A),\bar M_n)}{m^2}\\
 (|Z(A)|\leq M \textrm{ was used})\quad&\leq \bb E\left[\frac{(\bar M_n-m)^2}{m^2}\right]-\frac{\cov(\bar Z_n(A),\bar M_n)}{m^2}\\
&=\frac{\var(M)-\cov(Z_1(A),M_1)}{n\,m^2}\;.
\end{split}
\]
Hence, using the fact that $|\cov(Z_1(A),M_1)|^2\leq \var(M_1)\var(Z_1(A))\leq \var(M_1)\bb EM^2$, we obtain
\[
\begin{split}
\sup_A|\bb Q_n(A)-\bb Q(A)|&\leq \frac{\var(M)+\sup_A|\cov(Z_1(A),M_1)|}{n\,m^2}\\
&\leq \frac{\var(M)+\sqrt{\var(M)\bb EM^2}}{n\,m^2}
\end{split} 
\]
We can thus clearly see that the  convergence of $\bb Q_n(A)$ depends on the relative error of $M$. 

Next, we prove the bound for the mean absolute value. First, note that
the term $\bb E(\bar Z_n(A) \bar V_n)^2$, where $V_k:=M_k-m$, can be bounded  using the independence of the pairs $(Z_i(A),V_i)$ and $\bb EZ_i(A)=\bb E V_i=0$, as follows:
\[
\begin{split}
\bb E\bar Z_n^2(A)\bar V_n^2&=\frac{\sum_{i,j,k,l}
\bb E[Z_iZ_jV_kV_l]}{n^4}\leq \frac{3n^2-2n}{n^4}\max_{i,j,k,l}\bb E|Z_iZ_jV_kV_l|\leq \frac{3\bb E M^4}{n^2}\;.
\end{split}
\]
Therefore, using the triangle inequality,  we have:
\[
\begin{split}
 m\bb E|\widehat{\bb Q}_n(A)-\bb Q(A)|
&=\textstyle\bb E \left|\bar Z_n(A)-\frac{\bar Z_n(A)(\bar M_n-m)}{\bar M_n}\right|\\
( \bar M_n\geq 1)\qquad&\leq \bb E |\bar Z_n(A)|+\bb E|\bar Z_n(A)(\bar M_n-m)|\\
&\leq \sqrt{\bb E \bar Z_n^2(A)}+\sqrt{\bb E(\bar Z_n(A)(\bar M_n-m))^2}\\
 (|Z(A)|\leq M)\quad&\leq \frac{\sqrt{\bb E  M^2}}{\sqrt{n}}+\frac{\sqrt{3\bb E M^4}}{n}\;.
\end{split}
\]

\subsection{Proof of Theorem~\ref{thm:until t nonasymp}}
Recall that $N(t)=\widetilde N(t)+1$ is a stopping time.
Let $R(t):= T_{\widetilde N(t)+1}-t$, so that
 $r(t):=\bb ER(t)=m\bb E[N(t)] -t$. 
Using Wald's identity \eqref{Wald}, 
we can write:
\[
\begin{split}
\overline{\bb Q}_t(A)-\bb Q(A)
&=\textstyle\bb E\frac{\sum_{k=1}^{N(t)} (H_k(A)-M_k\bb Q(A))}{T_{N(t)}}\\
&=\textstyle\bb E\frac{\sum_{k=1}^{N(t)} Z_k(A)}{T_{N(t)}}= \bb E\frac{\frac{1}{t}\sum_{k=1}^{N(t)} Z_k(A)}{1+R(t)/t}\\
&=\textstyle\bb E\left(\frac{1}{1+R(t)/t}-1\right)\bar Z_t(A),
\end{split}
\]
where $\bar Z_t(A):=\frac{1}{t}\sum_{k=1}^{N(t)} Z_k(A)$.
Then,  using the fact that $\frac{1}{1+R(t)/t}\leq 1$,
we obtain the uniform bound
\[
\begin{split}
|\overline{\bb Q}_t(A)-\bb Q(A)|
&=\textstyle\frac{1}{t}\left|\bb E \left[\frac{R(t)}{1+R(t)/t}\bar Z_t(A)\right]\right|\\
&\leq \textstyle\frac{1}{t}\bb E\left|R(t)\bar Z_t(A)\right|\\
&\leq\frac{\sqrt{\bb E R^2(t) \bb E[\bar Z_t^2(A)]}}{t}\\
&=\frac{\sqrt{\bb ER^2(t)}}{t}\sqrt{\bb E[N(t)] \bb E[Z^2(A)/t^2]}\\
&=\frac{\sqrt{\bb E [R^2(t)]}}{t^{3/2}}\sqrt{\bb E[Z^2(A)]\bb E[N(t)]/t}\\
&\leq \frac{\sqrt{\bb E[R^2(t)]}}{t^{3/2}}\sqrt{\frac{\bb EM^2}{m}(1+r(t)/t)},
\end{split}
\]
where in the third  last line we used Wald's second-moment identity (see \eqref{Wald second-moment} below).
To finish the proof we apply Lorden's moment inequalities 
($\bb E[R(t)]\leq \bb E[M^2]/m$ and $\bb E[R^2(t)]\leq 4\bb E[ M^3]/(3m)$, see \cite{lorden1970excess}) to obtain
\begin{equation*}
\sup_A|\overline{\bb Q}_t(A)-\bb Q(A)|\leq 
 \frac{\sqrt{\frac{4}{3}\bb E[M^3]\bb E[M^2](m+\bb E[M^2]/t)}/m^3}{(t/m)^{3/2}}\;.
\end{equation*}

To prove the bound for the mean absolute value, we proceed as follows. Again using $\frac{1}{1+R(t)/t}\leq 1$, we have:
\[
\begin{split}
\bb E|\overline{\bb Q}_t(A)-\bb Q(A)|&=\textstyle\bb E\left|\frac{\frac{1}{t}\sum_{k=1}^{N(t)}Z_k(A)}{1+R(t)/t}\right|\\
&\leq \textstyle
\sqrt{\bb E\left(\frac{1}{t}\sum_{k=1}^{N(t)}Z_k(A)\right)^2}=\frac{\sqrt{\bb E[N(t)]\bb E Z^2(A)}}{t}\\
&\leq \sqrt{\frac{\bb E Z^2(A)}{tm}+\frac{\bb E[M^2]\bb E [Z^2(A)]}{t^2m^2}}\leq \frac{\sqrt{\bb E M^2}}{\sqrt{tm}}+\frac{\bb EM^2}{tm},
\end{split}
\]
where in the second last line we used Cauchy's inequality and Wald's second-moment identity, and
in the last line we used Lorden's inequality and the sub-additivity of the square root.

\subsection{Proof of Theorem~\ref{thm:until t}}

Denote $r(t):=\bb ER(t)$ and $r:=(\bb EM^2+m)/(2m)$ and note that under the condition 
$\bb EM^{p+5}<\infty$ for some $p\geq 0$, we have \citep{sGLY06a}
\[
 r(t)=r+o(1/t^{p+3}).
\]
Using 
$
0\leq\frac{1}{1+x}-1+x\leq x^2
$ for $x\geq 0$, we have the error bound:
\[
\begin{split}
|\overline{\bb Q}_t(A)-\bb Q(A)|&=\textstyle\left|\bb E\left(\frac{1}{1+R(t)/t}-1\right)\bar Z_t(A)\right|\\
(\textrm{triangle ineq.})\quad&\leq \textstyle \frac{|\bb E R(t)\bar Z_t(A)|}{t}+\left|\bb E\left(\frac{1}{1+R(t)/t}-1+\frac{R(t)}{t}\right)\bar Z_t(A)\right|\\
&\leq \textstyle\frac{|\bb E R(t)\bar Z_t(A)|}{t}+\frac{\bb ER^2(t)|\bar Z_t(A)|}{t^2}\\
&\leq\textstyle \frac{|\bb E R(t)\sum_{k=1}^{N(t)} Z_k(A)|}{t^2}+\frac{\sqrt{\bb E[R^4(t)]\bb E[\bar Z^2_t(A)]}}{t^2}\;.
\end{split}
\]
Since $\bb EM^{5}<\infty$, by Lorden's inequality, we have $\bb ER^4(t)<\infty$ and
the second term is $\c O(t^{-5/2})$, because  by Wald's second-moment identity:
\begin{equation}
\label{Wald second-moment}
\bb E[\bar Z_t^2(A)]=\frac{\bb E[N(t)]}{t^2}\bb E [Z_1^2(A)]\leq
(1+r(t)/t)\bb E [M^2]/t=\c O(1/t)\;.
\end{equation} 
For the first term, we verify that
 $e_A(t):=\bb E R(t)\sum_{k=1}^{N(t)}Z_k(A)<\infty$
satisfies the renewal equation  $e_A(t)=(u*v_A)(t)$ with
$
v_A(t):=\bb E [R(t)Z_1(A)]=\bb E [(R(t)-r)Z_1(A)],
$
see \cite[Page 25]{sAWA07a}.
The latter  is bounded uniformly in $A$: 
\[
\begin{split}
|v_A(t)|&=|\bb E [(R(t)-r)Z_1(A); M_1>t]+\bb E [(R(t)-r)Z_1(A); M_1\leq t]|\\
&=|\bb E [(M_1-r)Z_1(A); M_1>t]+\bb E [(r(t-M_1)-r)Z_1(A); M_1\leq t]|\\
&\leq \bb E [|M_1-r|M_1; M_1>t]+\bb E [|r(t-M_1)-r|M_1; M_1\leq t]\;.
\end{split}
\]
For the  first term, we obtain:
\[
\bb E [|M_1-r|M_1; M_1>t]=\c O(\bb E[M^{p+5};M>t]/t^{p+3})=o(1/t^{p+3})\;.
\]
For the second term, 
\[
\begin{split}
\bb E [|r(t-M)-r|M; M\leq t]&\leq 
\bb E [|r(t-M)-r|M; M\leq t/2]+\bb E [|r(t-M)-r|M; M\geq t/2]\\
&\leq \sup_{s>t/2}|r(s)-r|\bb E[M]+\sup_{s<t/2}|r(s)-r|\bb E[M;M>t/2]\\
&=o(1/t^{p+3})+o(1/t^{p+4})\;.
\end{split}
\]
Hence, we have the convergence uniformly in $A$:
\[
\begin{split}
e_A(t)&=\frac{\bb E(M_1-1-2r)M_1Z_1(A)}{2m}+o(1/t^{p+2})\\
&\leq\frac{\bb E|M_1-1-2r|M_1^2}{2m}+o(1/t^{p+2})\;.
\end{split}
\]
Putting it all together, we obtain
\[
\sup_A|\overline{\bb Q}_t(A)-\bb Q(A)|\leq\frac{\bb E|M-1-2r|M^2}{2m t^2} +\c O(t^{-5/2})+o(1/t^{p+4})\;.
\]
where $r=(\bb EM^2+m)/(2m)$. The exponential convergence comes from the fact that $\bb EM^p<\infty$ for all $p>0$, because $M\leq s^{\tau-1}$ is always bounded. This completes the proof.


\subsection*{Notational Setup  for Proofs of Theorems~\ref{theorem:a.s.} and \ref{theorem:simple} }

 We now introduce some  working notation that will apply to both the proofs of Theorem~\ref{theorem:a.s.} and \ref{theorem:simple}.  Define
\begin{equation}
\label{binary class}
\c F(\c Y_\cup)=\{ (b_1,\ldots,b_{T_n}) \in \{0,1\}^{T_n}: \textrm{  there exists an }  A\in \c A: b_i=\bb I{\{\v x_i\in  A\}},  \v x_i\in \c Y_\cup\}
\end{equation} 
to be a class of binary functions on $\{0,1\}^{T_n}$ such that each 
element of $\c F$ corresponds to an intersection 
of  $\c Y_\cup$ with a set $ A$ in $\c A$. Without any conditions on the class of sets $\c A$, the cardinality of $\c F(\c Y_\cup)$ grows exponentially in $T_n$, and we have
$|\c F(\c Y_\cup)|\leq 2^{T_n}$ for any $n$. 
Let 
\[
\c S_{\c A}(\{\v y_i\}_{i=1}^n)\idef\max_{\v y_1,\ldots,\v y_n}\#| A\cap\{\v y_1,\ldots,\v y_n\},  A\in \c A|
\]
denote the  \emph{Vapnik-Chervonenkis shatter coefficient} \citep{vapnik2013nature}. 
Loosely speaking, the shatter coefficient 
$\c S_{\c A}(\c Y_\cup)$ is the maximum number of distinct ways  in which the point-set $\c Y_\cup$
can intersect with elements of $\c A$.

  Sauer's Lemma \citep{sauer1972density} tells us that  if $\c A$ is a class of  sets with Vapnik-Chervonenkis dimension $v<\infty$, then the shatter  coefficient eventually grows polynomially in $n$, instead of exponentially: 
	\begin{equation}
	\label{lem:Sauer}
	\c S_{\c A}(\{\v y_i\}_{i=1}^n)\leq (n\m e/v)^v,\quad n>v\;.
	\end{equation}
Let $\rho_1,\ldots,\rho_n$ be iid random variables with marginal distribution $\bb P(\rho=\pm 1)=1/2$.
Let $\c Y_\cup'$ be a sample independent from $\c Y_\cup$ that can, in principle, be obtained from another $n$ independent calls to   Algorithm~\ref{algo:splitting}. The $\c Y_\cup'$ sample is a ``ghost''
sample \citep{gine1984some} that does not need to be constructed, but is only used in symmetrization inequalities.
We denote quantities computed using  $\c Y_\cup'$ by $H_i',\v Y_i',M_i',T_i'$, etc. For example, 
 $H'$ is an independent ``ghost'' copy of $H$.
We will make use of  two symmetrization inequalities by \cite{gine1984some}. 
The first will be used in Theorem~\ref{theorem:simple}:
\begin{equation}
\label{symmetrization E}
\bb E\sup_{ A\subseteq \c A}\left|\bar H_n( A)-\bb E H(A)\right|\leq\bb E\sup_{ A\subseteq \c A}\left| \frac{1}{n}\sum_{i=1}^n \rho_i (H_i(A)-H'_i(A)) \right|.
\end{equation}
The second will be used in Theorem~\ref{theorem:a.s.}: 
\begin{equation}
\label{symmetrization prob}
\bb P(\sup_{A\subseteq\c A}|H(A)-\bb E H(A)|>\epsilon)\leq 2\bb P(\sup_{A\subseteq\c A}|H(A)- H'(A)|>\epsilon/2)
\quad\textrm{for } \epsilon>\sqrt{8\sup_{A\subseteq\c A}\var( H(A))}.
\end{equation}

\subsection{Proof of Theorem~\ref{theorem:a.s.}}

If we can show that (with $g_n=o(n)\Leftrightarrow \lim_{n\uparrow\infty}g_n/n=0$), 
\begin{equation}
\label{exp bound}
\bb P(\sup_{A\subseteq \c A}|\widehat{\bb Q}_n(A)-\bb Q(\c A)|>\epsilon)\leq c_1\exp(-c_2 n\epsilon^2 +o(n))
\end{equation}
for some constants $c_1,c_2>0$,
then the fact that  $\sum_{n\geq 1} \bb P(\sup_{A\subseteq \c A}|\widehat{\bb Q}_n(A)-\bb Q(\c A)|>\epsilon)<\infty$ for any $\epsilon>0$ implies the almost sure convergence result of the theorem.  To show \eqref{exp bound} we will use the symmetrization inequality \eqref{symmetrization prob} and the simple union bound:
\begin{equation}
\label{simple bound}
\bb P(|X\pm Y|>\epsilon)\leq \bb P(|X|>\alpha\epsilon)+\bb P(|Y|>(1-\alpha)\epsilon)\qquad \alpha\in(0,1).
\end{equation}
 Using these two inequalities, we have
\[
\begin{split}
\bb P(\sup_{A\subseteq \c A}|\widehat{\bb Q}_n(A)-\bb Q(\c A)|>\epsilon)&\stackrel{\eqref{simple bound}}{\leq}\textstyle
\bb P(|\bar M_n-m|>\frac{m\epsilon}{2})+
\bb P(\sup_{A\subseteq \c A}|\bar H_n(A)-\bb E H(A)|>\frac{m\epsilon}{2}) \\
(\textrm{Hoeffding's with } M< s^{\tau} )\quad&\leq \textstyle
2\exp\left(-\frac{nm^2\epsilon^2}{2s^{2\tau} }\right)+
\bb P(\sup_{A\subseteq \c A}|\bar H_n(A)-\bb E H(A)|>\frac{m\epsilon}{2})\;. 
\end{split}
\]
Thus, in order to show \eqref{exp bound}, we   only need an exponentially decaying  bound
on the second term with $\epsilon_1=m\epsilon/2$:
\[
\begin{split}
\bb P(\sup_{A\subseteq \c A}|\bar H_n(A)-\bb E H(A)|>\epsilon_1) &\stackrel{\eqref{symmetrization prob}}{\leq}2\bb P(\sup_{A\subseteq\c A}|\bar H_n(A)- \bar H'_n(A)|>\epsilon_1/2)
\quad\textrm{for } \epsilon_1>\sqrt{8\frac{\bb EM^2}{n}}\Leftrightarrow
n>\frac{8\bb EM^2}{\epsilon_1^2}\;.
\end{split}
\] 
Recall that 
$\rho_1,\ldots,\rho_n$ is an iid random sample with  $\bb P(\rho=\pm 1)=1/2$, and that each $H_i'$ is an independent ``ghost'' copy of $H_i$. By symmetry, each 
$H_i(A)-H_i'(A)$ has the same distribution as 
$\rho_i(H_i(A)-H_i'(A))$. Using this observation, we obtain (with $\epsilon_2\idef \epsilon_1/2$ and for $n>2\bb EM^2/\epsilon^2_2$):
\[
\begin{split}
\bb P(\sup_{A\subseteq\c A}|\bar H_n(A)- \bar H'_n(A)|>\epsilon_2)
&=\bb P\left(\sup_{A\subseteq\c A}\frac{1}{n}\left|\sum_{i=1}^n\rho_i( H_i(A)-  H'_i(A))\right|>\epsilon_2\right)\\
&\stackrel{\eqref{simple bound}}{\leq}
\bb P\left(\sup_{A\subseteq\c A}\frac{1}{n}\left|\sum_{i=1}^n\rho_i H_i(A)\right|>\frac{\epsilon_2}{2}\right)+
\bb P\left(\sup_{A\subseteq\c A}\frac{1}{n}\left|\sum_{i=1}^n\rho_i H_i'(A)\right|>\frac{\epsilon_2}{2}\right)\\
&=2\bb P\left(\sup_{A\subseteq\c A}\frac{1}{n}\left|\sum_{i=1}^n\rho_i H_i(A)\right|>\frac{\epsilon_2}{2}\right)\;.
\end{split}
\]
The proof will be complete if we show that  ($\epsilon_3=\epsilon_2/2$)
\[\textstyle
\bb P\left(\sup_{A\subseteq\c A}\left|\frac{1}{n}\sum_{i=1}^n\rho_i H_i(A)\right|>\epsilon_3\right)\leq c_1\exp(-c_2 n\epsilon_3^2+o(n))
\]
for some constants $c_1,c_2>0$.
Let 
\[
N_{\c A}(\c Y_\cup)\idef \#| A\cap\{\v Y_1,\ldots,\v Y_{T_n}\},  A\in \c A|
\]
be the number of different subsets of the  points
in $\c Y_\cup$ that can be picked out by the class $\c A$
(so that, by definition, the shatter coefficient is 
 $\c S_{\c A}(T_n)=\max_{\c Y_\cup}N_{\c A}(\c Y_\cup)$).
Similarly, let 
\[
N_{\c A}(\overline{\c Y}_\cup)\idef \#| A\cap\overline{\c Y}_\cup,  A\in \c A|,
\]
where $\overline{\c Y}_\cup=\{\v Y_1,\ldots,\v Y_{n s^{\tau-1}}\}$ is the collection of all $ns^{\tau-1}$ potential states from $n$ independent runs of splitting \citep{vLEC18s}[Section 3.1] (in practice only a small fractions of  these trajectories survive till the final level of splitting). Clearly, $N_{\c A}(\overline{\c Y}_\cup)\geq N_{\c A}(\c Y_\cup)$.

A well-known result (see \citep{rao1962relations} and \citep[Theorem 13.13]{devroye2013probabilistic}) asserts that when the $\v Y$'s have a density and  $\c A$ is the class of all convex sets, then:
\begin{equation}
\label{devroye}
\bb E N_{\c A}(\overline{\c Y}_\cup)=2^{o(n)}\;.
\end{equation}
Thus, by conditioning on $\c Y_\cup$, we can write:
\[
\begin{split}
\bb P\left(\sup_{A\subseteq\c A}\left|\frac{1}{n}\sum_{i=1}^n\rho_i H_i(A)\right|>\epsilon_3\right)&=\bb E\left[\bb P\left(\sup_{A\subseteq\c A}\left|\frac{1}{n}\sum_{i=1}^n\rho_i H_i(A)\right|>\epsilon_3\,\Bigg |\,\c Y_\cup\right)\right]\\
\textrm{(union bound)}\quad&\leq \bb E\left[N_{\c A}(\c Y_\cup) \sup_{A\subseteq\c A}\bb P\left(\left|\frac{1}{n}\sum_{i=1}^n\rho_i H_i(A)\right|>\epsilon_3\,\Bigg |\,\c Y_\cup\right)\right]\\
(\textrm{Hoeffding's with } |\rho_i H_i(A)|< s^{\tau} )\quad&\leq \bb E\left[N_{\c A}(\c Y_\cup) \sup_{A\subseteq\c A}2\exp\left(-\frac{2n\epsilon_3^2}{(2s^\tau)^2}\right)\right]\\
(\textrm{using } |\c Y_\cup|\leq|\overline{\c Y}_\cup|=ns^{\tau-1})\quad&\leq 2\exp\left(-\frac{n\epsilon_3^2}{2s^{2\tau}}\right)\bb E\left[N_{\c A}(\overline{\c Y}_\cup)\right]\\
&\stackrel{\eqref{devroye}}{=}2\exp\left(-\frac{n\epsilon_3^2}{2s^{2\tau}}+o(n)\right)\;.
\end{split}
\]
This completes the proof.

\subsection{Proof of Theorem~\ref{theorem:simple}}

\label{append:theorem 1}
Our proof follows as closely as possible the proof of the classical VC inequalities, as described in \cite[Theorems 3.1 \& 3.2]{devroyecombinatorial}.

Applying  the triangle inequality and then the symmetrization inequality \eqref{symmetrization E}, yields:
\[
\begin{split}
\bb E\sup_{ A\subseteq \c A}|\widehat{\bb Q}_n( A)-\bb Q( A)|&\leq
\bb E\sup_{ A\subseteq \c A}\left|\frac{\bar H_n( A)}{\bar M_n}-\frac{\bar H_n( A)}{m}\right|+\bb E\sup_{ A\subseteq \c A}\left|\frac{\bar H_n( A)}{m}-\bb Q(A)\right|\\
&\leq \frac{1}{m}\bb E|\bar M_n-m|+\frac{1}{m\, n}\bb E\sup_{ A\subseteq \c A}\left| \sum_{i=1}^n \rho_i (H_i(A)-H'_i(A)) \right|\\
&=\frac{\sqrt{\var(M)}}{m\sqrt{n}}+\frac{1}{m\, n}\bb E\Psi(\{\v Y_k\},\{\v Y_k'\}),
\end{split}
\]
where we define the conditional expectation
\[
\Psi(\{\v Y_k\},\{\v Y_k'\})\idef\bb E\left[\sup_{A\subseteq \c A}\left| \sum_{i=1}^n \rho_i (H_i( A)-H'_i(A)) \right|\Bigg|\c Y_\cup,\c Y_\cup'\right],
\]
and the last expectation is with respect to  $\v\rho$.
Let $\breve{\c A}\subset \c A$ be the collection of sets such that all intersections with the pointset 
$\{\v Y_1,\ldots,\v Y_{T_n}, \v Y_1',\ldots,\v Y_{T_n}'\}$ are represented once, and 
any two sets in
$\breve{\c A}$ are different. Observe that
\begin{equation*}
\label{conditioning trick}
\Psi(\{\v Y_k\},\{\v Y_k'\})=\bb E\left[\sup_{A\subseteq \breve{\c A}}\left| \sum_{i=1}^n \rho_i (H_i(A)-H'_i(A)) \right|\Bigg|\c Y_\cup,\c Y_\cup'\right]
\end{equation*}
and that
$|\breve{\c A}|\leq \c S_{\c A}(T_n+T_n')$.

Let $\|X\|_G<\infty$ denote the sub-Gaussian coefficient of the random variable $X$. In other words, the moment generating function of $X$ satisfies
\[
\bb E\exp(t X)\leq \exp(t^2\|X\|_G^2/2),\quad\forall t\;.
\]
We shall next use the maximal inequality
\begin{equation}
\label{maximal}
\bb E\max_{k\in \s K}|X_k|\leq\sqrt{2\ln(2|\s K|)}\max_{k\in \s K}\|X_k\|_G
\end{equation}
 for a finite index set $\s K$, which holds even if the  $X_k$'s are dependent.
We will also make use of the  property that  
\begin{equation}
\label{subGaussian}\textstyle
\|\sum_k w_k X_k\|_G^2=\sum_k w_k^2\|X_k\|_G^2,
\end{equation}
 whenever $X_1,X_2,\ldots$ are independent.
Conditioning on all $\{\v Y_k\},\{\v Y_k'\}$, and taking expectation over  $\v\rho$, we obtain:
\[
\begin{split}
\Psi(\{\v Y_k\},\{\v Y_k'\})&=\bb E_{\v\rho}\sup_{A\subseteq \breve{\c A}}\left| \sum_{k=1}^{n} \rho_k (H_k(A)-H_k'(A))  \right|\\
(\textrm{maximal ineq.})\quad&\stackrel{\eqref{maximal}}{\leq}
\sqrt{2\ln(2\s S_{\c A}(T_n+T_n'))}\sup_{A\subseteq \c A}\left\|\sum_{k=1}^n \rho_k (H_k(A)-H_k'(A))\right\|_G\\
(\textrm{Sauer's Lemma})\quad&\stackrel{\eqref{lem:Sauer}+\eqref{subGaussian}}{\leq} \sqrt{2\ln(2[(T_n+T_n')\m e/v]^v)}\sup_{A\subseteq \c A}\sqrt{
\sum_{k=1}^n\|\rho_k (H_k(A)-H_k'(A))\|_G^2}\\
&\leq \sqrt{2\ln(2[(T_n+T_n')\m e/v]^v)}\sup_{A\subseteq \c A}\sqrt{
\sum_{k=1}^n  (H_k(A)\vee H_k'(A))^2 }\\
&\leq  \sqrt{2\ln(2[(T_n+T_n')\m e/v]^v)  \sum_{k=1}^n (M_i\vee M_i')^2}\;.
\end{split}
\]
Therefore, using the bound ($R_i\idef(M_i\vee M_i')$, $r_n^2\idef\frac{1}{n}\sum_iR_i^2$):
\[
\begin{split}
\textstyle\bb E r_n^2\ln(T_n+T_n')&\leq \textstyle
\ln(2n)\bb E r_n^2 +\bb E\left(\frac{1}{n}\sum_iR_i^2\right)\ln(\frac{1}{n}\sum_i R_i)  \\
\textrm{(Cauchy-Schwartz)}\quad&\leq\textstyle
\ln(2n)\bb E r_n^2 +\bb E\left(\frac{1}{n}\sum_iR_i^2\right)\frac{1}{2}\ln(\frac{1}{n}\sum_i R_i^2)  \\
\textrm{(Jensen's on $x\ln(x)$)}\quad&\textstyle\leq \ln(2n)\bb E r_n^2+ \bb E\frac{1}{2n}\sum_i R_i^2\ln( R_i^2)\\
&= \ln(2n)\bb E R^2+ \bb E R^2\ln( R)\\
&\leq 2\ln(2n)\bb E M^2+ 2\bb E M^2\ln( M),
\end{split}
\]
we obtain:
\[
\begin{split}
\bb E\sup_{A\subseteq \c A}|\widehat{\bb Q}_n(A)-\bb Q(A)|&\textstyle\leq
\frac{\sqrt{\var(M)}}{m\sqrt{n}}+\frac{\bb E\sqrt{2(\ln(2)+v\ln(T_n+T_n')+v-v\ln(v))r_n^2}}{m\sqrt{n}}\\
&\leq \textstyle\frac{\sqrt{\var(M)}}{m\sqrt{n}}+\frac{2\sqrt{(\ln(2)+v+v\ln (2n/v))\bb EM^2+v\bb EM^2\ln M}}{m\sqrt{n}}\\
&\leq \textstyle\frac{\sqrt{\var(M)}}{m\sqrt{n}}+\frac{2\sqrt{v\,\ln(2n)\,\bb E[M^2\ln M]}}{m\sqrt{n}}\;\psi_1(v,n),
\end{split}
\]
where
\[
\psi_1(v,n)=\sqrt{\frac{(\ln(2)+v+v\ln (2n/v))\bb EM^2}{v\ln(2n)\bb EM^2\ln M}+\frac{1}{\ln(2n)}}\;.
\]
This completes the proof of the theorem.

\subsection{Proof of Theorem~\ref{theorem:hard}}

We need to introduce more working notation.  First, recall a number of standard definitions.
Define the weighted $L_p(\bb P)$ metric on the probability space $(\bb R^d,\s B,\bb P)$ via the norm
$
\|X\|_{p}:=
\left(\int_{\bb R^d}| X(\omega)|^p\di\bb P(\omega)\right)^{1/p},\; p\geq 1
$.
Let $\c F$ be a class of functions. 
An $\epsilon$-cover 
 of $\c F$ under the $L_p(\bb P)$ metric   is a finite set  $\c C=\{f_1,\ldots,f_{ |\c C|}\}$ with cardinality $|\c C|$ such that  for every $f\in \c F$ there exists an $f_k\in \c C$ that satisfies
$\|f-f_k\|_{p}\leq\epsilon
$. Let $\c C^*$ be the $\epsilon$-cover with the smallest cardinality.  The  cardinality of
	the smallest  $\epsilon$-cover of $\c F$ under the metric  $L_p(\bb P)$ is called the \emph{covering number} and is denoted by
  $\s N(\epsilon,\c F,L_p(\bb P))$. We will write 
	$\s N(\epsilon,\c F,L_p(\bb P))=\s N(\epsilon,\c F)$ if the metric is clear from the context.

Recall that $\c Y_\cup$ with $T_n=n\bar M_n=|\c Y_\cup|$ is the agglomeration of all the final states from $n$ independent runs of  Algorithm~\ref{algo:splitting}. Since  the splitting factor is $s$, we have  $M\leq s^{\tau}$. Denote $\sigma_n^2:=\frac{1}{n}\sum_{j=1}^n M_j^2$. We know that $\sigma_n\leq s^{\tau}$. For each index
$k=0,1\ldots,K:=\lceil\log_s(s^{\tau}\sqrt{n})\rceil$, we define a cover  as follows. 

Conditional on $\c Y_\cup$, we let $\c C^*_k$ be the smallest $(\sigma_n s^{-k})$-cover of the set of  functions 
\[
\c F(\c Y_\cup)=\{ (H_1(A),\ldots,H_n(A));  A\in \c A\},
\] under the weighted metric with norm 
$
 \|\v h\|_2:=\sqrt{\frac{1}{n}\sum_{j=1}^{n}h_j^2} 
$.

Observe that the zero vector  is within $\sigma_ns^{-0}$ radius of all elements of $\c F(\c Y_\cup)$, and  that $\c C_0=\{\v 0\}$  is an minimal $(\sigma_ns^{-0})$-cover, that is, $\s N( \sigma_ns^{-0},\c F(\c Y_\cup))=1$.
Further, the minimal $\epsilon$-cover for 
$\epsilon\in [0,\frac{1}{\sqrt{n}})$ contains all the elements of $\c F(\c Y_\cup)$, that is, $\s N( \sigma_ns^{-K},\c F(\c Y_\cup))=|\c F(\c Y_\cup)|=|\c C_K^*|$.

Conditional on $\c Y_\cup$, we let $\v h=(h_1,\ldots,h_{n})$ be the vector with components $h_j(A)=\sum_{k\in \c Y_j}\bb I{\{\v Y_k\in A\}}$ (each $h_j$ is a conditional version of $H_j$).
 For a given $\v\rho=(\rho_1,\ldots,\rho_{n})^\top$, let $\v h^*$
correspond to the vector maximizing 
\[
\sup_{A\subseteq\c A}\left|\sum_{k=1}^{n}\rho_k h_k( A)\right|
=\left|\sum_{k=1}^{n}\rho_k h_k^*\right|=|\v\rho^\top\v h^*|\;.
\]
Then, for $k=0,\ldots,K$, let $\v h_k$ be the vector in the minimal cover $\c C^*_k$, which is closest to $\v h^*$, that is
$
\|\v h_k-\v h^*\|_2= \inf_{\v h\in \c C^*_k}\|\v h-\v h^*\|_2\leq \sigma_ns^{-k}
$.
It follows that 
$
\v h^*=\v h_K=\sum_{k=1}^{K}(\v h_k-\v h_{k-1})
$.
By the triangle inequality we have
\[
\|\v h_k-\v h_{k-1}\|_2\leq \|\v h_k-\v h^*\|_2+\|\v h_{k-1}-\v h^*\|_2
\leq (s+1)\sigma_n s^{-k}\;.
\]
Hence, 
\begin{equation}
\label{proof:bound}
\begin{split}
|\v\rho^\top\v h^*|&=\left|\sum_{k=1}^{K}\v\rho^\top(\v h_k-\v h_{k-1})\right|\leq \sum_{k=1}^{K}\left|\v\rho^\top(\v h_k-\v h_{k-1})\right|\\
&\leq \sum_{k=1}^{K}\max_{\substack{\v h\in \c C_k^*,\v h'\in \c C_{k-1}^*\\\|\v h-\v h'\|_2<(s+1) \sigma_n s^{-k}}}\left|\v\rho^\top(\v h-\v h')\right|
\end{split}
\end{equation}
Taking expectation with respect to $\v\rho$ and using the maximal inequality \eqref{maximal}, we thus obtain
\[
\bb E_{\v\rho}\max_{\substack{\v h\in \c C_k^*,\v h'\in \c C_{k-1}^*\\\|\v h-\v h'\|_2<(s+1) \sigma_n s^{-k}}}\left|\v\rho^\top(\v h-\v h')\right|\leq \sqrt{2\ln(2|\c C_{k-1}^*||\c C_k^*|)}\; (s+1) \sigma_n s^{-k} \;.
\]
Therefore, taking expectation over $\c Y_\cup$:
\[
\bb E|\v\rho^\top\v h^*|\leq (s+1)\sum_{k=1}^{K} s^{-k} \bb E\left[\sigma_n\sqrt{2\ln(2|\c C_k^*|^2)}  \right] \leq (s+1)\sum_{k=1}^{K} s^{-k}\sqrt{2\bb E\left[\ln(2|\c C_k^*|^2)\sigma_n^2\right]} \;.
\]
Finally, from the triangle inequality and symmetrization inequality \eqref{symmetrization E}, we have
\[
\begin{split}
\bb E\sup_{A\subseteq \c A}|\widehat{\bb Q}_n(A)-\bb Q(A)|&\leq
\bb E\sup_{A\subseteq \c A}\left|\frac{\bar H_n(A)}{\bar M_n}-\frac{\bar H_n(A)}{m}\right|+\bb E\sup_{A\subseteq \c A}\left|\frac{\bar H_n(A)}{m}-\bb Q(A)\right|\\
&\leq \frac{1}{m}\bb E|\bar M_n-m|+\frac{2}{m\, n}\bb E\sup_{\c A\subseteq \c B}\left| \sum_{i=1}^n \rho_i H_i(A)  \right|\\
&\leq  \frac{\sqrt{\var(M)}}{m\sqrt{n}}+\frac{(s+1)2\sqrt{2}}{m\, n}
\sum_{k=1}^{K} s^{-k}\textstyle\sqrt{\bb E\left[\sigma_n^2\ln(2\s N^2(\sigma_ns^{-k},\c F(\c Y_\cup)))\right]}\;.
\end{split}
\]
It thus remains to bound  the metric entropy $\ln \s N(\sigma_n\epsilon,\c F(\c Y_\cup))$. 
For a fixed $\c Y_\cup$, let $\c C'_1,\ldots,\c C'_n$ be minimal $\epsilon$-covers corresponding to each of the $n$
binary function classes ($j=1,\ldots,n$):
\[
\c F(\c Y_j)=\{ (b_1,\ldots,b_{M_j}):   A\in \c A, b_i=\bb I{\{\v Y_i\in A\}},  \v Y_i\in \c Y_j\}\;.
\]
 This implies that for any $\v b_j\in \c F(\c Y_j)$, there exists
	an $\v s_j\in \c C_j'$ such that:
	\[
	\|\v b_j-\v s _j\|_2=\sqrt{\frac{1}{M_j}\sum_{k=1}^{M_j} (b_j^{(k)}-s_j^{(k)})^2}\leq \epsilon\;.
	\]
Then, the set 
$
\left\{s^{(1)}_j+\cdots+s^{(M_j)}_j: \v s_j\in \c C'_j, j=1,\ldots,n\right\}
$
is an $\sigma_n\epsilon$-cover of $\c F(\c Y_\cup)$. To see this, note that for any $\v h\in
\c F(\c Y_\cup)$,  we have 
\[
h_j\in\left\{b_{j}^{(1)}+\cdots+b_{j}^{(M_j)}: \v b_j \in \c F(\c Y_j)\right\},\quad j=1,\ldots,n
\]
 and by the Cauchy-Schwartz inequality:
\[
\begin{split}\textstyle
\left\|\v h-\left(\sum_{k=1}^{M_1}s_1^{(k)},\ldots,\sum_{k=1}^{M_n}s_n^{(k)}\right)\right\|_2^2&=\textstyle\frac{1}{n}\sum_{j=1}^n \left(h_j-\sum_{k=1}^{M_j}s_j^{(k)}\right)^2\\
&=\textstyle\frac{1}{n}\sum_{j=1}^n M_j^2 \left(\frac{1}{M_j}\sum_{k=1}^{M_j}(b_j^{(k)}-s_j^{(k)})\right)^2 \\
& \leq \textstyle
\frac{1}{n}\sum_{j=1}^n M_j^2 \left(\frac{1}{M_j}\sum_{k=1}^{M_j}|b_j^{(k)}-s_j^{(k)}|\right)^2\\
&\leq \textstyle
\frac{1}{n}\sum_{j=1}^n M_j^2 \left(\frac{1}{M_j}\sum_{k=1}^{M_j}|b_j^{(k)}-s_j^{(k)}|^2\right)\leq \sigma_n^2 \epsilon^2\;.
\end{split}
\]
Using  the inequality of \cite{haussler1995sphere} 
\begin{equation}
\label{haussler}
\ln\c N(\epsilon,\c F(\c Y_j))\leq \ln(\m e[v+1])+v\ln(2\m e/\epsilon^2), \qquad \epsilon\in[0,1]
\end{equation}
for the
cover number of a class of sets $\c A$ with VC dimension $v<\infty$, 
we thus have the bound on the metric entropy of $\c F(\c Y_\cup)$:
\[
\begin{split}
\ln\c N(\sigma_n\epsilon,\c F(\c Y_\cup))&\leq\sum_{j=1}^n \ln\c N(\epsilon,\c F(\c Y_j))\stackrel{\eqref{haussler}}{\leq} n(\ln(\m e[v+1])+v\ln(2\m e/\epsilon^2)),\qquad \epsilon\in[0,1]\;.
\end{split}
\]
Hence, combining all the results so far we obtain the upper bound for
 $\bb E\sup_{A\subseteq \c A}|\widehat{\bb Q}_n(A)-\bb Q(A)|-\frac{\sqrt{\var(M)}}{m\sqrt{n}}$:
\[
\begin{split}
\frac{(s+1)2\sqrt{2}}{m\, n}
\sum_{k=1}^{K} \frac{\sqrt{\bb E\left[\sigma_n^2\ln(2\c N^2(\sigma_ns^{-k},\c F(\c Y_\cup)))\right]}}{s^k}&\leq 
\frac{(s+1)4\sqrt{v\bb EM^2}}{m\sqrt{n}}\sum_{k=1}^{K} \frac{\sqrt{\frac{\ln(2)}{2nv}+\frac{\ln(\m e[v+1])}{v}+\ln(2\m es^{2k})}}{s^k}\\
&\leq \frac{(s+1)4\sqrt{v\bb EM^2}}{m\sqrt{n}}\psi_2(\tau,v,n)\;.
\end{split}
\]
Hence, the result of the theorem follows.

\end{APPENDICES}


\end{document}